\documentclass[a4paper,11pt]{article}

\usepackage{jheppub} 

\usepackage[T1]{fontenc} 
\usepackage{inputenc}

\usepackage{amsmath,bbm,array,amsfonts,graphicx,wrapfig,arydshln,lscape,float,multirow,longtable,rotating,makecell}
\usepackage{url}

\DeclareMathAlphabet\mathbfcal{OMS}{cmsy}{b}{n}

\newcommand{\la}{\langle}
\newcommand{\ra}{\rangle}

\newcommand{\eq}{\begin{equation}}

\newcommand{\eqe}{\end{equation}}
\newcommand{\eqa}{\begin{eqnarray}}
\newcommand{\eqae}{\end{eqnarray}}

\newcommand{\f}{\frac}
\newcommand{\bea}{\begin{eqnarray}}
\newcommand{\eea}{\end{eqnarray}}
\newcommand{\beas}{\begin{eqnarray*}}
\newcommand{\eeas}{\end{eqnarray*}}
\newcommand{\ndt}{\noindent}
\renewcommand{\b}{\mathbf}
\newcommand{\mo}{\mathcal{O}}
\newcommand{\nn}{\nonumber}

\usepackage{tikz}
\usepackage[compat=1.1.0]{tikz-feynman}
\usepackage{subfigure}
\usepackage{pdfpages}
\usepackage{scalerel}

\title{\boldmath On-Shell Electroweak Sector and the Higgs Mechanism}

\author[a]{Brad Bachu}
\author[a]{Akshay Yelleshpur}
\affiliation[a]{Department of Physics, Princeton University, NJ, USA 08540}

\emailAdd{bbachu@princeton.edu}
\emailAdd{ysakshay@princeton.edu}

\abstract{We take the first steps towards an entirely on-shell description of the bosonic electroweak sector of the Standard Model. We write down on-shell three particle amplitudes consistent with Poincare' invariance and little group covariance. Tree-level, four particle amplitudes are determined by demanding consistent factorization on all poles and correct UV behaviour. We present expressions for these $2\rightarrow 2$ scattering amplitudes using massive spinor helicity variables. We show that on-shell consistency conditions suffice to derive relations between the masses of the $W^\pm, Z$, the Weinberg angle and the couplings. This provides a completely on-shell description of the Higgs mechanism without any reference to the vacuum expectation value of the Higgs field.}

\begin{document} 
\maketitle
\flushbottom

\section{Introduction}
Quantum fields, path integrals and Lagrangians have been a cornerstone of 20th century theoretical physics. They have been used to describe a variety of natural phenomena accurately. Yet, it is becoming increasingly apparent that these mathematical tools are both inefficient and insufficient. They obscure the presence of a deeper, underlying structure, particularly of scattering amplitudes in quantum field theories. The field of scattering amplitudes has undergone a paradigm shift in the past three decades. This was sparked by the discovery of the stunning simplicity of the tree-level gluon scattering amplitudes in \citep{parke1, parke2}. The simplicity of these amplitudes was revealed due to the use of helicity spinors, $(\lambda_\alpha, \tilde{\lambda}_{\dot{\alpha}})$ which correspond to the physical degrees of freedom of massless particles - helicity. The forbidding complexity of the Feynman diagram based calculation of tree level gluon scattering amplitudes is now understood to be an artefact of the unphysical degrees of freedom introduced by gauge redundancy. These unphysical degrees of freedom are necessary to package the physical degrees of freedom into local quantum fields in a manner consistent with Poincare' invariance \cite{weinberg}. The simplicity of these amplitudes fueled the development of a variety of ``on-shell" techniques for computing scattering amplitudes involving massless particles. These methods do not rely on Feynman diagrams, do not suffer from gauge redundancies and do not invoke virtual particles. For an overview of these methods, see \cite{review1, review2, review3, review4, review5} and the references therein.  However, most of this progress was limited to amplitudes involving only massless particles. \\

\ndt Since helicity spinors correspond to the physical degrees of freedom of massless particles, it is natural to attempt to find variables akin to these for massive particles. The physical degrees of freedom of massive particles correspond to the little group $SU(2)$ \cite{wigner}. Some early generalizations can be found in \cite{othermassive1, othermassive2, othermassive3, othermassive4, othermassive5, othermassive6, othermassive7, othermassive8}. However, the little group  covariance was not manifest in these generalizations until the introduction of Spin-spinors (or massive spinor-helicity variables) in \cite{massivesh}. These variables $(\lambda_{\alpha}^I, \tilde{\lambda}_{\dot{\alpha}}^I)$  which carry both little group indices and Lorentz indices and make the little group structure of amplitudes manifest. Information about all the $(2S+1)$ spin components of each particle is packaged into compact, manifestly Lorentz invariant expressions. Amplitudes written in terms of these variables are directly relevant to physics. This is in contrast to a Feynman diagram based computation which involves an intermediate object with Lorentz indices which must then be contracted with polarization tensors which carry the little group indices. For some interesting applications of these variables, ranging from black holes to supersymmetric theories see \cite{applications1, applications2, applications3, applications4, applications5, applications6}.\\
 
\ndt One of the biggest successes of path integrals and the Lagrangian formulation is the development of effective field theory and the Higgs mechanism. Recently, efforts have been made towards the development of effective field theory using on-shell methods \cite{eft1, eft2, eft3, eft4, eft5, eft6}. A completely on-shell description of the Higgs mechanism was outlined in \cite{massivesh} for the abelian and non-abelian gauge theories. The conventional understanding of the Higgs mechanism involves a scalar field acquiring a vacuum expectation value and vector bosons becoming massive by ``eating'' the goldstone modes arising from spontaneously broken symmetry. However, the on-shell description has no mention of scalar fields, potentials and vacuum expectation values. Nevertheless, it reproduces all of the same physics. Additionally, well known results like the Goldstone Boson equivalence theorem become trivial consequences of the high energy limits of our expressions. From an amplitudes perspective, it is more natural to think of the Higgs mechanism as a unification of the massless amplitudes in the UV into massive amplitudes in the IR. In this paper, we will focus on computing scattering amplitudes in the bosonic electroweak sector of the standard model and describing the Higgs mechanism and electroweak symmetry breaking using a completely on-shell language.\\

\ndt The paper is structured as follows. We begin with a brief review of the little group, spin-spinors and their properties in Section~[\ref{sec:review}]. We focus on constructing three particle amplitudes in the IR in Section~[\ref{sec:IR}] and the UV in Section~[\ref{sec:UV}]. In Section~[\ref{sec:matching}], we compute the high energy limits of the three particle amplitudes in the IR and demand that they are consistent with the three point amplitudes in the UV. This gives us the all the standard relations between the coupling constants, the masses of the Z and $W^\pm$ and the Weinberg angle $\theta_w$. We also see the emergence of the custodial SO(3) symmetry in the limit in which the hypercharge coupling vanishes. Finally, in Section~[\ref{sec:4ptamplitudes}], we construct 4 point amplitudes in the IR by gluing together the three point amplitudes found before. We enunciate the details involved in the gluing process. We will also discover that demanding that these amplitudes have a well defined high energy limit imposes constraints on the structure of the theory.

\section{Scattering amplitudes and the little group}\label{sec:review}

\subsection{Helicity spinors and spin-spinors}
In this section, we briefly review some aspects of the on-shell approach to constructing scattering amplitudes. We will review the formalism of spin-spinors introduced in \cite{massivesh} whilst highlighting some features important for this paper. One particle states are irreducible representations of the Poincare' group. They are labeled by their momentum and a representation of the little group. If the particle is charged under any global symmetry group, appropriate labels must be appended to these. In (3+1) spacetime dimensions, the little groups for massless and massive particles are $SO(2)$ and $SO(3)$ respectively. \\

\ndt Representations of the massless little group, $SO(2)=U(1)$ can be specified by an integer corresponding to the helicity of the massless particle. A massless one particle state is thus specified by its momentum and helicity. Under a Lorentz transformation $\Lambda$, 
\bea
\label{eq:masslesstransform}
|p, h, \sigma \rangle \rightarrow w^{-2h} |\Lambda p, h, \sigma \rangle \, ,
\eea  
where $\sigma$ are labels of any global symmetry group and $w$ has the same meaning as in \cite{weinberg} and \cite{massivesh}. It is useful to introduce elementary objects $\lambda_\alpha,\tilde{\lambda}_{\dot{\alpha}}$  which transform under the little group as 
\bea
\label{eq:masslesslambdatransform}
\lambda_\alpha \rightarrow w^{-1} \lambda_\alpha \qquad\text{and}\qquad  \tilde{\lambda}_{\dot{\alpha}} \rightarrow w  \tilde{\lambda}_{\dot{\alpha}} \, .
\eea
We can use these objects to build representations with any value of $h$. The natural candidates for these elementary objects are the spinors which decompose the null momentum $p_{\alpha \dot{\alpha}} \equiv p_\mu \sigma^\mu_{\alpha \dot{\alpha}}$. We have
\bea
\label{eq:lambdadef}
p_{\alpha \dot{\alpha}} =  \lambda_\alpha \tilde{\lambda}_{\dot{\alpha}} \equiv  |\lambda\ra_\alpha[\tilde\lambda|_{\dot\alpha} \,.
\eea
Throughout the paper we will find it convenient to make use of the following notation,
\beas
\lambda_\alpha \equiv | \lambda \rangle \qquad \tilde{\lambda}_{\dot{\alpha}} \equiv [\tilde\lambda| \qquad \lambda^\alpha \equiv \la\lambda | \qquad  \tilde{\lambda}^{\dot{\alpha}} \equiv |\tilde\lambda] \, .
\eeas 
For any two null momenta $p_1, p_2$, we can form two Lorentz invariant combinations of these spinors, 
\bea
\label{eq:masslessinv}
\langle 12 \rangle \equiv \epsilon^{\alpha \beta} (\lambda_1)_\beta (\lambda_2)_\alpha \qquad \qquad [12] \equiv \epsilon^{\dot{\alpha} \dot{\beta}} (\tilde{\lambda}_1)_{\dot{\alpha}}(\tilde{\lambda}_2)_{\dot{\beta}} .
\eea

\ndt The massive little group is $SO(3)=SU(2)$. It representations are well known and can be specified by the value of the Casimir operator which is restricted to values $S(S+1)$, where $S$ is defined as the spin of the particle. The spin S representation is $2S+1$ dimensional. A massive one-particle state of spin S thus transforms as a tensor of rank 2S under $SU(2)$.
\bea
\label{eq:massivetransform}
|p, I_1,\dots I_{2S}, \, \sigma \rangle \rightarrow W_{I_1J_1}\dots W_{I_{2S}J_{2S}} |\Lambda p, J_1,\dots J_{2S},\, \sigma \rangle
\eea
The elementary objects in this case are the spinors of $SU(2)$. These transform as 
\bea
\label{eq:massivelambdatransform}
{\boldsymbol \lambda}_\alpha^I \rightarrow \left(W^{-1}\right)^{I}_{J}  {\boldsymbol \lambda}_\alpha^J \qquad  \tilde{\boldsymbol\lambda}_\alpha^I \rightarrow W^{I}_{J}  \,\,\tilde{\boldsymbol\lambda}_\alpha^J 
\eea
Higher representations can be built by taking tensor products of these. A decomposing the rank 2 momentum, similar to eq.(\ref{eq:lambdadef}) yields the requisite spinors. 
\beas
p_{\alpha \dot{\alpha}} =  \epsilon_{JI}|\boldsymbol{\lambda} \rangle^I [{\boldsymbol{\tilde{\lambda}}}|^J = \epsilon_{JI}\tilde{\boldsymbol\lambda}_\alpha^I \tilde{\boldsymbol\lambda}_{\dot{\beta}}^J,
\eeas
Note that we have det$(p) = $ det$(\lambda)$ det$(\tilde{\lambda}) = m^2$. For the rest of the paper, we will set det$(\lambda) = $ det$(\tilde{\lambda}) = m$\footnote{There is more freedom to set $\det\lambda = M$ and $\det\tilde\lambda = \tilde M$ such that $M\tilde M = m^2$, but for our purposes $M = \tilde M = m$ suffices.}. We will find it convenient to suppress the little group indices on the spin-spinors. We do this according to the convention in eq. (\ref{eq:massivesuppress}). Finally, we can construct Lorentz invariants out of spin-spinors corresponding to two massive momenta $p_1, p_2$ similar to eq.(\ref{eq:masslessinv}).
\bea
\label{eq:massiveinv}
\langle \b 1 \b 2 \rangle^{IJ} \equiv \epsilon^{\alpha \beta} (\boldsymbol\lambda_1)_\beta^I (\boldsymbol\lambda_2)_\alpha^J \qquad \qquad [\b 1 \b 2]^{IJ} \equiv \epsilon^{\dot{\alpha} \dot{\beta}} (\tilde{\boldsymbol\lambda}_1)_{\dot{\alpha}}^I(\tilde{\boldsymbol\lambda}_2)_{\dot{\beta}}^J .
\eea
\subsection{Scattering amplitudes as little group tensors}
Scattering amplitudes are defined as the overlap of in and out states. We have
\beas
\mathcal{M}(p_1, \rho_1 \dots p_n, \rho_n) = {}_{\text{out}}\langle p_1, \rho_1, \dots p_n, \rho_n | 0\rangle_{\text{in}} 
\eeas
where we are assuming that all particles are outgoing. $\rho = (h, \sigma)$ for massless particles (eq.(\ref{eq:masslesstransform})) and $\rho = (\left\lbrace I_1, \dots I_{2S} \right\rbrace, \sigma)$ for massive ones (eq.(\ref{eq:massivetransform})). Translation invariance allows us to pull out a delta function which imposes momentum conservation
\bea
\label{eq:momconservation}
\mathcal{M}(p_1, \rho_1 \dots p_n, \rho_n)  = \delta^4 (p_1+ \dots p_n) M(p_1, \rho_1, \dots, p_n, \rho_n)
\eea 
Assuming that the asymptotic multi-particle states transform under Lorentz transformations as the tensor products of one-particle states, we have the following transformation law for the function $M(p_1, \rho_1, \dots, p_n, \rho_n)$ under a Lorentz transformation $\Lambda$.
\bea
\label{eq:amplitudetransform}
M(p_a, \rho_a) \rightarrow \prod_a \left( D_{\rho_a \rho'_a}(W)\right)  M((\Lambda p)_a, \rho'_a) 
\eea
where $D_{\rho_a \rho'_a}(W) = \delta_{\sigma, \sigma'_a}\delta_{h_a, h'_a} w^{-2h_a}$ and $D_{\rho_a \rho'_a}(W) = \delta_{\sigma, \sigma'_a}W^{I_1}_{I'_1} \dots W^{I_{2S}}_{I'_{2S}}$. As an example, we display the transformation law for a 4-particle amplitude where particle 1 is massive with spin 1, particle 2 is massless with helicity $5/2$, particle 3 is massless with
helicity $-2$ and particle 4 is massive with spin 0.
\begin{equation*}
   M^{\{I_1,I_2\},\{5/2\},\{-2\},\{0\}}(p_1,p_2,p_3,p_4) \rightarrow (W_1)^{I_1}_{I'_1}\,\,(W_1)^{I_2}_{I'_2}\,\,w_2^{-5}\,\,w_3^{4} \,\,\, M^{\{I'_1,I'_2\},\{5/2\},\{-2\},\{0\}}(p_1,p_2,p_3,p_4)
\end{equation*}
Thus, objects constructed from helicity spinors and spin spinors can correspond to scattering amplitudes only if they are Lorentz invariant and have the above transformation law under the little group. This imposes restrictions on the functional forms objects that make up scattering amplitudes. Indeed, three point amplitudes involving all massless particles are completely fixed by this restriction. At three points, we have, 
\beas
2p_1.p_2 = \langle 12\rangle [12] = 0 \qquad 2p_2.p_3 = \langle 23\rangle [23] = 0 \qquad 2p_3.p_1 = \langle 31 \rangle [31] = 0
\eeas 
We must choose either the $\lambda$ or the $\tilde{\lambda}$ to be proportional to each other. The two solutions are the MHV configuration
\bea
\label{MHV}
\tilde{\lambda}_1 = \langle 23\rangle \tilde{\zeta} \qquad \tilde{\lambda}_2 = \langle 31\rangle \tilde{\zeta} \qquad \tilde{\lambda}_3 = \langle 12\rangle \tilde{\zeta}  
\eea 
and the anti-MHV configuration
\bea
\label{MHV}
\lambda_1 = [23] \zeta \qquad \lambda_2 = [31] \zeta \qquad \lambda_3 = [12] \zeta 
\eea 
Using these, one can show that the three point amplitudes can only take the following form.
\begin{alignat}{3}
\label{eq:3particlemassless}
    M^{h_1 h_2 h_3} &= g\, \langle 12\rangle^{h_1+h_2-h_3} \langle 23\rangle^{h_2+h_3-h_1} \langle 31\rangle^{h_3+h_1-h_2} \,,\quad &\text{ if}\quad h_1+h_2+h_2 >0 \nonumber\\
 & =\tilde{g}\, [12]^{h_3-h_1-h_2} [23]^{h_1-h_2-h_3} [31]^{h_2-h_3-h_1}\,,\qquad &\text{ if}\quad h_1+h_2+h_2 >0
\end{alignat}
In cases involving one or more massive particles, Lorentz invariance and little group covariance are not sufficient to completely fix the amplitude. However, they narrow down the form of the amplitude to a finite number of terms. For an exhaustive analysis, we refer the reader to \cite{massivesh} and \citep{csm1, csm2}. In this paper, we will discuss only the amplitudes relevant to us.
\subsection{The high energy limit of spin-spinors}
When particle energies are much higher than their masses, it is intuitive to treat them as massless. We can formalize this by expanding the spin-spinors in a convenient basis in little group space.
\begin{alignat}{4}\label{eq:spinorexpansion}
\lambda_\alpha^I &=& \lambda_\alpha \zeta^{-I} + \eta_\alpha \zeta^{+I} &=&\sqrt{E + p}\, \zeta_\alpha^{+}(p) \,\zeta^{-I}(k) + \sqrt{E-p}\, \zeta_\alpha^-(p)\,\zeta^{+I}(k)\nonumber \\
\tilde{\lambda}_{\dot{\alpha}I} &=& \tilde{\lambda}_{\dot{\alpha}}\zeta_I^+ - \tilde{\eta}_{\dot{\alpha}}\zeta_I^- &=&  \sqrt{E+p}\, \tilde{\zeta}_\alpha^-(p)\, \zeta_I^+(k) - \sqrt{E-p}\, \tilde{\zeta}_{\dot{\alpha}}(p)\, \zeta_I^-(k)
\end{alignat}
where $\lambda, \tilde{\lambda}$ are the helicity spinors, $\zeta^{\pm I}$ are eigenstates of spin 1/2 along the momentum. We give explicit expressions for all objects involved are in Appendix~[\ref{sec:conventions}]. Here, we just note that $\eta_\alpha, \tilde{\eta}_{\dot{\alpha}} \propto \sqrt{E-m} = m + \mathcal{O}(m^2)$. Taking the high energy limit corresponds to taking $m/E \rightarrow 0$. In this limit, the spin-spinors reduce to massless helicity ones. Finally, it should be pointed out that we must take special care while taking the high energy limit of 3-point amplitudes. Owing to the special three point kinematics, factors like $\langle \b 1 \b 2\rangle$ or $[\b 1 \b 2]$ can tend to zero in the high energy limit.

\section{Three particle amplitudes}
3-particle amplitudes the fundamental building blocks of scattering amplitudes. In this paper, we are interested in analyzing the bosonic content of the standard model both in the UV and IR. The spectrum in the UV is comprised solely of massless particles. The three point amplitudes are completely determined by Poincare' invariance and little group scaling as outlined in Section~[\ref{sec:review}]. The form of these amplitudes was given in eq.(\ref{eq:3particlemassless}). We will use this formula to write down all the relevant 3-particle amplitudes in Section~[\ref{sec:UV}]. All the amplitudes in the UV obey the $SU(2)_L \times U(1)_Y$ symmetry.\\

\ndt The spectrum in the IR consists of massive particles and a single massless vector. 3-particle amplitudes involving massive particles aren't completely fixed. They can have several contributing structures. In Section~[\ref{sec:IR}], we will write down all the relevant amplitudes. The amplitudes in the IR obey only a $U(1)_{\text{EM}}$ symmetry. \\

\ndt Finally, we will demand that the high energy limit of the IR amplitudes is consistent with the amplitudes in the UV. We will find that this consistency is possible only if the masses of particles in the IR are related in a specific way. These turn out to be the usual relations involving the Weinberg angle.  
\subsection{The IR}\label{sec:IR}
The spectrum in the IR consists of the following particles. 
\begin{itemize}
      \item Three massive spin 1 bosons $\left( W^+, W^-, Z\right)$ which have masses $(m_{\scaleto{W}{4pt}}, m_{\scaleto{W}{4pt}}, m_{\scaleto{Z}{4pt}})$ and charges $(+, -, 0)$ respectively under a global symmetry group $U(1)_{EM}$. Note that $W^+$ and $W^-$ are eigenstates of the $U(1)_{\text{EM}}$ generator. They must have equal mass as they are related by charge conjugation. 
      \item One massless spin 1 boson, the photon, $\gamma$ which is not charged under the $U(1)_{EM}$.
      \item One massive scalar, the higgs, $h$ which is also uncharged under $U(1)_{EM}$.  
\end{itemize} 
The only symmetry of the IR is the $U(1)_{\text{EM}}$. We will now discuss all the relevant three point amplitudes in the IR. Owing to the existence of various identities amongst the spin-spinors, each amplitude can be written in a multitude of different ways. In many of the cases below, we have chosen particularly convenient ways of writing them. Different form three point amplitudes lead to different expressions for four point amplitudes. The difference between these are contact terms that can be fixed by imposing other constraints on the amplitude. While the form of the contact term will depend on the form of the three point amplitudes used, the final amplitude will be the same. We will elaborate on these comments in the appropriate places below.\\ 
\makebox[\linewidth]{\rule{\textwidth}{1pt}}
\begin{center}
$\b{W^+ W^- Z}$
\end{center} 
\begin{equation} \label{eq:WWZ}
\vcenter{\hbox{\includegraphics[scale = 1]{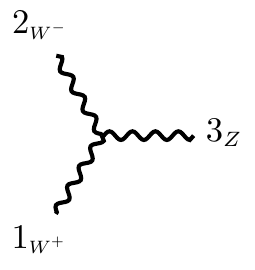}}} = \frac{e_{\scaleto{W}{4pt}} }{m_{\scaleto{W}{4pt}}^2 m_{\scaleto{Z}{4pt}}} \left[\la\b{12}\ra [\b{12}]\la\b{3}|p_1 - p_2|\b{3}] + \text{cyc.} \right]
\end{equation}
This is a form of the three point amplitude that is chosen to suit our needs. It should be noted that it can be reduced to a combination of $<\,>$ and $[\,\,]$. As an example, consider the first term in the above equation which can be re-written as follows.
\beas
\b{ \langle 12 \rangle [12] \langle 3|}p_1-p_2|\b{3\rangle} = 2\left( m_1 \b{[12]\langle 23\rangle [31]} - m_2 \b{ [12][23] \langle 31\rangle} + m_3 \b{[12] \langle 23\rangle \langle 31\rangle} \right)
\eeas
where we made use of the Sch$\ddot{\text{o}}$uten identity $\b{ \langle 12\rangle 3 + \langle 23\rangle 1 + \langle 31 \rangle 2 } = 0.$\\
\makebox[\linewidth]{\rule{\textwidth}{1pt}}	
\begin{center}
$\b{W^+ W^- \gamma}$
\end{center} 
\begin{equation} \label{eq:WWA}
\vcenter{\hbox{\includegraphics[scale=1]{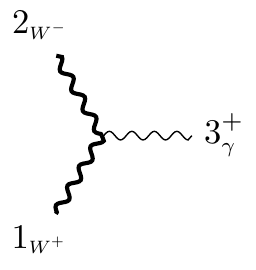}}} = e\,\, x_{12}^+\b{\la12\ra^2}
\end{equation}
We discuss other forms of writing the same vertex in Appendix [\ref{sec:mincoupling}].\\
\makebox[\linewidth]{\rule{\textwidth}{1pt}}
\begin{center}
$\b{Z Z h}$
\end{center} 
\begin{equation}\label{eq:ZZH}
  \vcenter{\hbox{\includegraphics[scale=1]{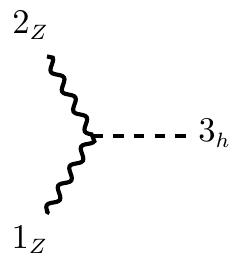}}} = \frac{e_{HZZ}}{m_{\scaleto{Z}{4pt}}} \la \b{12}\ra[\b{12}]  + \frac{\mathcal{N}_1}{m_{\scaleto{Z}{4pt}}} \left( \la\b{12}\ra^2 + [\b{12}]^2 \right)
\end{equation}
\makebox[\linewidth]{\rule{\textwidth}{1pt}}
\newpage
$$\b{W^+ W^- h}$$
\begin{equation}\label{eq:WWH}
  \vcenter{\hbox{\includegraphics[scale=1]{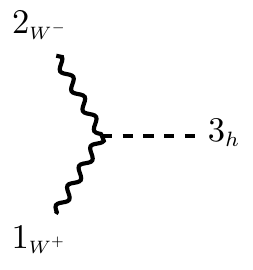}}} = \frac{e_{\scaleto{WWH}{4pt}}}{m_{\scaleto{W}{4pt}}} \la \b{12}\ra[\b{12}] + \frac{\mathcal{N}_2}{m_{\scaleto{W}{4pt}}} \left( \la\b{12}\ra^2 + [\b{12}]^2 \right)
\end{equation}
We will set $\mathcal{N}_1 = \mathcal{N}_2 = 0$ in what follows as yield four point amplitudes which grow as $E^2$ where $E$ is the center-of-mass energy. 
\subsection{The UV}\label{sec:UV}
The UV spectrum of the electroweak sector of the standard model consists of the following
\begin{itemize}
    \item One massless spin-0 particle $\Phi = \{\phi_1, \phi_2,\phi_3,\phi_4\}$ with four real degrees of freedom in the fundamental representation of $\mathrm{SO(4)} = \mathrm{SU(2)}_L \times \mathrm{SU(2)}_R$.
    \item One massless spin-1 particle  $B$ with charge $\frac{1}{2}$ under a global $\mathrm{U}_Y(1)$  symmetry group.
    \item Three massless spin-1 particles $(W_1, W_2, W_3)$, in the adjoint representation of $\mathrm{SU}(2)_L$. These are not charged under the group $U(1)_Y$. In order to facilitate easy comparison to the massive particles in the IR, we will work with particle states $W^\pm = \frac{1}{\sqrt{2}}\left( W^1 \pm i W^2\right)$ which are eigenstates of the $U(1)_{\text{EM}}$ symmetry in the IR.
\end{itemize}
The electroweak sector has an $SU(2)_L \times U(1)_Y$ symmetry. The generators of these symmetries are related to the generators of $SO(4)$ listed in Appendix~[\ref{sec:grouptheory}] as follows.
\bea
\label{eq:UVgenerators}
T^{1} \equiv X^{1} \qquad T^2 \equiv X^2 \qquad T^3 = X^3 \qquad T^B = Y^3 \,.
\eea
The generator of $U(1)_{\text{EM}}$, which we denote by $Q$ can be written as a linear combination of the generators of the UV. 
\bea
\label{eq:EMgenerator}
e\, Q = \alpha \, g\,  T^3 + \beta \, g'\,  T^B
\eea
where $e$ is $U(1)_{\text{EM}}$ coupling. Since  $T^{\pm} = \frac{1}{\sqrt{2}}\left( T^1 \pm i T^2\right) $ are eigenstates of $Q$, 
we are free to work with the states $W^{\pm}$ in the UV. We will now list all the relevant amplitudes in the UV. The superscripts on the particles indicate the corresponding helicities.\\
\makebox[\linewidth]{\rule{\textwidth}{1pt}}

$$\b{W^+ W^- W^3}$$

\begin{equation}
\label{eq:WWW3}
\vcenter{\hbox{\includegraphics[scale=1]{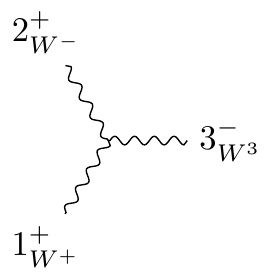}}}
    = g\, \frac{\la 1 2\ra^3}{\la23\ra\la31\ra}
\end{equation}
\makebox[\linewidth]{\rule{\textwidth}{1pt}}
\begin{center}
$\b{W^+ \Phi \Phi}$
\end{center} 
\begin{equation}
\label{eq:Wpff}
\vcenter{\hbox{\includegraphics[scale=1]{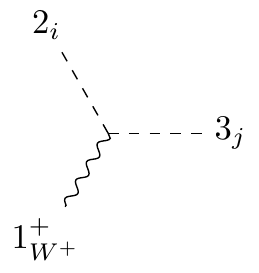}}}
    = g(T^+)_{ij} \frac{\la 12 \ra \la31\ra}{\la23\ra}
\end{equation}
\makebox[\linewidth]{\rule{\textwidth}{1pt}}
\begin{center}
$\b{W^- \Phi \Phi}$
\end{center} 
\begin{equation}
\label{eq:Wmff}
\vcenter{\hbox{\includegraphics[scale=1]{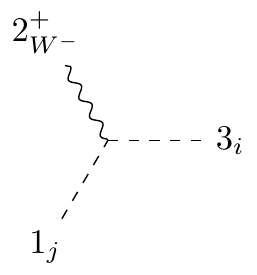}}}
    = g(T^-)_{ij} \frac{\la 12 \ra \la23\ra}{\la31\ra}
\end{equation}
\makebox[\linewidth]{\rule{\textwidth}{1pt}}
\begin{center}
$\b{W^3 \Phi \Phi}$
\end{center} 
\begin{equation}
\label{eq:W3ff}
\vcenter{\hbox{\includegraphics[scale=1]{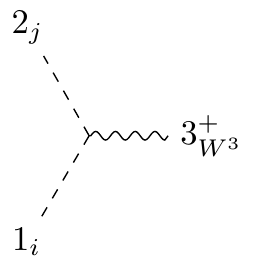}}}
    = g(T^3)_{ij} \frac{\la 23 \ra \la31\ra}{\la12\ra}
\end{equation}
These amplitudes must be proportional to a generator $T^{\{+,-,3\}}$ of the SU(2). For explicit forms of these generators, see Appendix~[\ref{sec:grouptheory}].\\
\makebox[\linewidth]{\rule{\textwidth}{1pt}}
\begin{center}
$\b{B \Phi \Phi}$
\end{center} 
\begin{equation}
\label{eq:Bff}
\vcenter{\hbox{\includegraphics[scale=1]{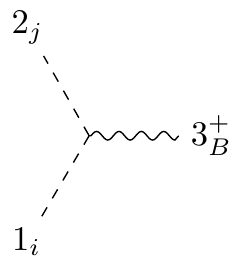}}}
    = g^\prime(T^B)_{ij} \frac{\la 23 \ra \la31\ra}{\la12\ra}
\end{equation}
\makebox[\linewidth]{\rule{\textwidth}{1pt}}
The above list doesn't contain any amplitudes which involve $W's$ and $B$ since the $W$'s are not charged under the $U(1)_Y$.
Note that all the above amplitudes involve particles whose helicities, $h_1, h_2, h_3$ are such that $\sum h_i >0$. The amplitudes with $\sum h_i <0$ are given by flipping $<\,> \rightarrow [\,\,]$.
 
\subsection{The HE Limit of the IR}
All the amplitudes in the IR listed above have one or more factors of $\frac{1}{m}$. At first glance, this seems to suggest that they blow up in the UV and cannot be matched onto any 3-particle amplitude of massless particles. However, we will see that all these factors of inverse mass drop out when we take the special 3 particle kinematics into account and carefully take the high energy limit. Many of these high energy limits are worked out in \cite{massivesh} and \cite{csm1}. We present them here in a form compatible with our conventions. For each massive leg, in order to take the high energy limit we must first specify the component which we are interested in.\\
\makebox[\linewidth]{\rule{\textwidth}{1pt}}
\begin{center}
$\b{W^+ W^- Z}$
\end{center}
\bea
\label{eq:WWZ:HE:Components}
\vcenter{\hbox{\includegraphics[scale = 1]{figs/WWZ.pdf}}} \underrightarrow{\quad\text{HE}\quad}
\begin{cases}
   \vcenter{\hbox{\includegraphics[scale = 1]{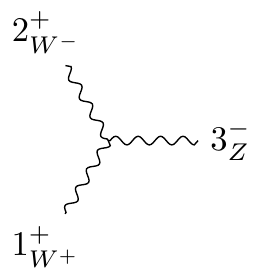}}}
    &=  e_{\scaleto{W}{4pt}} \frac{\la12\ra^3}{\la23\ra\la31\ra} \\
   \vcenter{\hbox{\includegraphics[scale = 1]{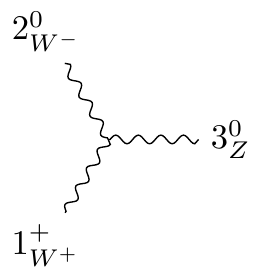}}}
    &=-e_{\scaleto{W}{4pt}}\f{m_{\scaleto{Z}{4pt}}}{m_{\scaleto{W}{4pt}}} \frac{\la12\ra\la31\ra}{\la23\ra} \\
   \vcenter{\hbox{\includegraphics[scale = 1]{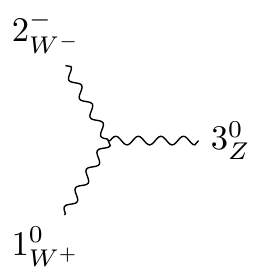}}}
    &= -e_{\scaleto{W}{4pt}}\f{m_{\scaleto{Z}{4pt}}}{m_{\scaleto{W}{4pt}}}\frac{\la12\ra\la23\ra}{\la31\ra}  \\
   \vcenter{\hbox{\includegraphics[scale = 1]{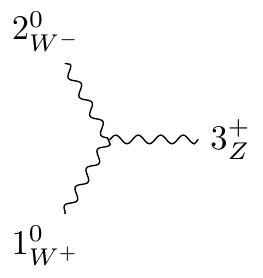}}}
    &= e_{\scaleto{W}{4pt}} \f{m_{\scaleto{Z}{4pt}}^2-2m_{\scaleto{W}{4pt}}^2}{m_{\scaleto{W}{4pt}}^2} \frac{\la23\ra\la31\ra}{\la12\ra}
\end{cases} 
\eea
Amplitudes with one longitudinal mode and two transverse vanish in the high energy limit.\\
\makebox[\linewidth]{\rule{\textwidth}{1pt}}
\newpage
\begin{center}
$\b{W^+ W^- \gamma}$
\end{center}
\bea
\label{eq:WWA:HE:Components}
\vcenter{\hbox{\includegraphics[scale=1]{figs/WWA.pdf}}} \underrightarrow{\quad\text{HE}\quad}
\begin{cases}
   \vcenter{\hbox{\includegraphics[scale = 1]{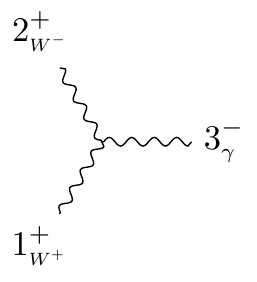}}}
    &= e \frac{\la23\ra^3}{\la23\ra\la31\ra}  \\
   \vcenter{\hbox{\includegraphics[scale = 1]{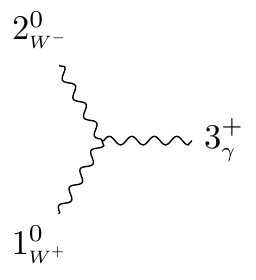}}}
    &= -2e\frac{\la23\ra\la31\ra}{\la12\ra}
\end{cases}
\eea
\makebox[\linewidth]{\rule{\textwidth}{1pt}}
\begin{center}
$\b{W^+ W^- h} \text{ and } \b{Z Z h}$
\end{center}
\bea
\label{eq:ZZH-WWH:HE:Components}
\vcenter{\hbox{\includegraphics[scale=1]{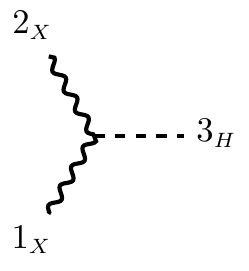}}} \underrightarrow{\quad\text{HE}\quad}
\begin{cases}
\vcenter{\hbox{\includegraphics[scale=1]{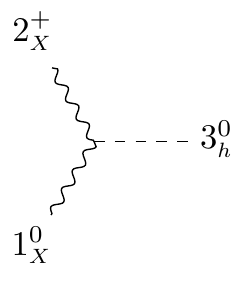}}}
 &=    -\f{e_{HXX} }{2} \frac{\la12\ra\la23\ra}{\la31\ra}   \\
\vcenter{\hbox{\includegraphics[scale=1]{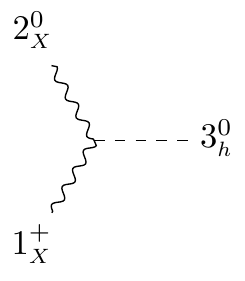}}}
 &=   \f{e_{HXX}}{2} \frac{\la12\ra\la31\ra}{\la23\ra}
\end{cases}
\eea
where $X = W,Z$. Amplitudes involving only one transverse mode vanish in the high energy limit.

\subsection{UV-IR consistency}\label{sec:matching}
Thus far, we have specified the structure of the IR which consists of the interactions among the $W^\pm,Z, \gamma$ and $h$ which preserve the $U(1)_{\text{EM}}$ symmetry and the structure of the UV which consists of the interactions among the $W^a, B, \Phi$ which preserve the $SU(2)_L \times U(1)_Y$ symmetry. We must now ensure that they are compatible with each other. We take the high energy limit of the IR amplitudes and demand that they are equal to the appropriate amplitudes in the UV. We refer to this process as `UV-IR matching'. This imposes many constraints and determines the couplings in the IR in terms of those in the UV. Furthermore, it also imposes constraints on the masses of the particles in the IR. To begin with, we must relate the degrees of freedom in the IR to the ones in the UV. We assume that they are related by the following orthogonal transformation 
\bea
\label{eq:UVIRrotation}
\begin{pmatrix}
W^+\\
W^-\\
Z\\
\gamma
\end{pmatrix} = 
\begin{pmatrix}
\mathcal{O}_{++} & \mathcal{O}_{+-} & \mathcal{O}_{+3} & \mathcal{O}_{+B} \\
\mathcal{O}_{-+} & \mathcal{O}_{--} & \mathcal{O}_{-3} & \mathcal{O}_{-B} \\
\mathcal{O}_{Z+} & \mathcal{O}_{Z-} & \mathcal{O}_{Z3} & \mathcal{O}_{ZB} \\
\mathcal{O}_{\gamma +} & \mathcal{O}_{\gamma -} & \mathcal{O}_{\gamma3 } & \mathcal{O}_{\gamma B}  
\end{pmatrix}
\begin{pmatrix}
W^+\\
W^-\\
W^3\\
B
\end{pmatrix}
\eea
Clearly, we must have $\mathcal{O}_{+-} = \mathcal{O}_{+3} =\mathcal{O}_{+B} = \mathcal{O}_{-+} = \mathcal{O}_{-3} = \mathcal{O}_{-B} = 0$. This is a result of working with the same states in the UV and IR. Orthogonality demands that the matrix be block diagonal, and so we have the simpler relation  
\bea
\begin{pmatrix}
Z\\
\gamma
\end{pmatrix} = \begin{pmatrix}
\text{cos }\theta_w & -\text{sin }\theta_w \\
\text{sin }\theta_w & \text{cos }\theta_w
\end{pmatrix} 
\begin{pmatrix}
W^3\\
B
\end{pmatrix}
\eea
for some unknown angle $\theta_w$. All the massive particles in the IR have longitudinal components which must be generated by some some linear combination of the scalars in the UV. We assume that
\bea
W^{+^{(0)}} = U_{W^+i}\Phi_i \qquad W^{-^{(0)}} = U_{W^-i}\Phi_i \qquad Z^{^{(0)}} = U_{Zi}\Phi_i
\eea
The remaining linear combination of the components of $\Phi$,  $h = U_{hi}\Phi_i$ has an independent existence. Indeed, it is well known that its presence is crucial for the theory to have a good UV behaviour. The high energy limit of each of the three point amplitudes in the IR must be equal to some combination of the amplitudes in the UV. This determines the masses in the IR in terms of the couplings in the UV. It also imposes some constraints on the couplings in the UV. All the constraints arising from eq.(\ref{eq:WWZ:HE:Components}) - eq.(\ref{eq:ZZH-WWH:HE:Components}) are determined below.
\makebox[\linewidth]{\rule{\textwidth}{1pt}}

$$\b{W^+ W^- Z}$$

There are a total of 27 components to the $W^+W^-Z$ amplitude corresponding to the $(+,-,0)$ spin component of each particle. Amplitudes with just one longitudinal mode all vanish in the high energy limit. This is consistent with the fact that there are no $WW\Phi,\, WB\Phi,\, BB\Phi $ amplitudes in the UV. The independent constraints arising from the remaining components are given below. Recall that the superscript on the particle is its helicity. These are also listed at the top of each diagram.\\
\makebox[\linewidth]{\rule{10cm}{.25pt}}
\begin{center}
$(++-)$
\end{center}
\begin{align}
   \vcenter{\hbox{\includegraphics[scale=1]{figs/Wp+Wm+Z-.pdf}}} \nonumber
   &\equiv \mathcal{O}_{Z3}  \vcenter{\hbox{\includegraphics[scale=1]{figs/WpWmW3.pdf}}} 
\end{align}
Using the expressions from eq.(\ref{eq:WWZ:HE:Components}) and eq.(\ref{eq:WWW3}), we get
\bea
      e_{\scaleto{W}{4pt}} \frac{\la12\ra^3}{\la23\ra\la31\ra}  =  g\,\mathcal{O}_{Z3}\,\frac{\la 1 2\ra^3}{\la23\ra\la31\ra} 
\implies e_{\scaleto{W}{4pt}} = g\, \text{cos }\theta_w \, .
\eea
The absence of a $W^+W^-B$ interaction in the UV means that there is no term proportional to $\mo_{ZB}$ on the RHS.]\\
\makebox[\linewidth]{\rule{10cm}{.25pt}}
$$(0\,0\,+)$$

\begin{align}
   \vcenter{\hbox{\includegraphics[scale=1]{figs/Wp0Wm0Z+.pdf}}} & \equiv U_{W^+i}\,U_{W^-j}\,\mathcal{O}_{Z3} \vcenter{\hbox{\includegraphics[scale=1]{figs/PhiPhiW3.pdf}}} + U_{W^+i}\,U_{W^-j}\,\mathcal{O}_{ZB}\vcenter{\hbox{\includegraphics[scale=1]{figs/PhiPhiB.pdf}}} \nonumber \, .
\end{align}

Using eq.(\ref{eq:WWZ:HE:Components}), eq.(\ref{eq:W3ff}) and eq.(\ref{eq:Bff}) in the above gives,
\begin{align}
   e_{\scaleto{W}{4pt}} \f{m_{\scaleto{Z}{4pt}}^2-2m_{\scaleto{W}{4pt}}^2}{m_{\scaleto{W}{4pt}}^2} \frac{\la23\ra\la31\ra}{\la12\ra} &= U_{W^+i}\,\left(g \,\mathcal{O}_{Z3}\,T^3_{ij} + g^\prime\, \mathcal{O}_{ZB}\, T^B_{ij} \right)U_{W^-j}\frac{\la23\ra\la31\ra}{\la12\ra}  \nonumber \\
   \implies  e_{\scaleto{W}{4pt}} \f{m_{\scaleto{Z}{4pt}}^2-2m_{\scaleto{W}{4pt}}^2}{m_{\scaleto{W}{4pt}}^2} &= U_{W^+i}\left(g\, \text{cos }\theta_w \, T^3_{ij} - g^\prime\, \text{sin }\theta_w\, T^B_{ij}  \right) U_{W^-j} \,.
\end{align}
\makebox[\linewidth]{\rule{10cm}{.25pt}}
\begin{center}
$(+\,0\,0)$
\end{center}
\begin{align}
   \vcenter{\hbox{\includegraphics[scale=1]{figs/Wp+Wm0Z0.pdf}}} &=\, g\, U_{W^-i} \, U_{Zj} \vcenter{\hbox{\includegraphics[scale=1]{figs/WpPhiPhi.pdf}}} \nonumber
\end{align}
Again, eq.(\ref{eq:WWZ:HE:Components}) and eq.(\ref{eq:Wpff}) give
\begin{align}
   -e_{\scaleto{W}{4pt}}\frac{m_{\scaleto{Z}{4pt}}}{m_{\scaleto{W}{4pt}}} \frac{\la 12 \ra \la31\ra}{\la23\ra} &=  \, g\, U_{W^-i} \, T^+_{ij}\, U_{Zj}  \,\frac{\la 12 \ra \la31\ra}{\la23\ra}\nonumber \\
   \implies  -e_{\scaleto{W}{4pt}}\frac{m_{\scaleto{Z}{4pt}}}{m_{\scaleto{W}{4pt}}} &=  \, g\, U_{W^-i}\, T^+_{ij}\, U_{Zj} \,. 
\end{align}
\makebox[\linewidth]{\rule{\textwidth}{1pt}}
\begin{center}
$\b{W^+W^-\gamma}$
\end{center}
Since the photon is massless in the IR, the $W^+W^-\gamma$ amplitude only has 18 components. This leads to the following constraints.
\newpage
\makebox[\linewidth]{\rule{10cm}{.25pt}}\\

$$(++-)$$
\begin{align}
  \vcenter{\hbox{\includegraphics[scale=1]{figs/Wp+Wm+A-.pdf}}} &\equiv \mathcal{O}_{\gamma 3} \vcenter{\hbox{\includegraphics[scale=1]{figs/WpWmW3.pdf}}}\nonumber
\end{align}

\begin{align}
     e_{\scaleto{W}{4pt}} \frac{\la12\ra^3}{\la23\ra\la31\ra}  &= g\, \mathcal{O}_{\gamma \, 3}  \frac{\la 1 2\ra^3}{\la23\ra\la31\ra} \nonumber \\
   \implies e &= g\, \text{sin }\theta_w
\end{align}

\makebox[\linewidth]{\rule{10cm}{.25pt}}

$$(0\,0\,+)$$
\begin{align}
   \vcenter{\hbox{\includegraphics[scale=1]{figs/Wp0Wm0A+.pdf}}} &\equiv U_{W^+i}\,U_{W^-j}\,\mathcal{O}_{\gamma3}\vcenter{\hbox{\includegraphics[scale=1]{figs/PhiPhiW3.pdf}}}
   + U_{W^+i}\,U_{W^-j}\,\mathcal{O}_{\gamma B}\vcenter{\hbox{\includegraphics[scale=1]{figs/PhiPhiB.pdf}}} \nonumber
\end{align} 

\begin{align}
 -2e\frac{\la23\ra\la31\ra}{\la12\ra} &=  U_{W^+i}\,\left(g\, \mathcal{O}_{A3} \, T^3_{ij} \,+\, g^\prime\, \mathcal{O}_{AB} \, T^B_{ij} \, \right)U_{W^-j}\frac{\la23\ra\la31\ra}{\la12\ra} \nonumber \\
   \implies -2e &= U_{W^+i}\left( g\,\text{sin }\theta_w T^3_{ij} + g^\prime \,\text{cos }\theta_w\, T^B_{ij}\right)U_{W^-j}
\end{align}

\makebox[\linewidth]{\rule{\textwidth}{1pt}}
$$\b{W^+Z\gamma}$$
Conservation of the $U(1)_{\text{EM}}$ charge in the IR must be imposed. This is achieved by setting the $W^+Z\gamma$ amplitude to zero. A similar equation is given by setting the $W^-Z\gamma$ amplitude to zero.\\
\makebox[\linewidth]{\rule{10cm}{.25pt}}
$$(0\, 0\, +)$$

\begin{align}
   \vcenter{\hbox{\includegraphics[scale=1]{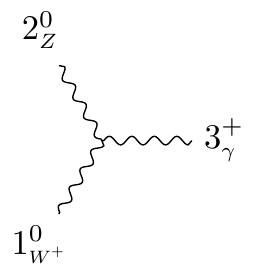}}} &\equiv U_{W^+i}\,U_{Zj}\,\mathcal{O}_{\gamma 3}\vcenter{\hbox{\includegraphics[scale=1]{figs/PhiPhiW3.pdf}}}
   + U_{W^+i}\,U_{Zj}\,\mathcal{O}_{\gamma B}\vcenter{\hbox{\includegraphics[scale=1]{figs/PhiPhiB.pdf}}} \nonumber
\end{align}
\begin{align}
   0 &= U_{W^+i}\left(g\,\mathcal{O}_{\gamma 3} \, T^3_{ij} \, + \, g^\prime \,\mathcal{O}_{\gamma B} \, T^B_{ij}\right)U_{Zj}\frac{\la23\ra\la31\ra}{\la12\ra} \nonumber \\
   \implies 0 &=  U_{W^+i}\left(g \, \text{sin }\theta_w \, T^3_{ij} \, + \, g^\prime \, \text{cos }\theta_w \, T^B_{ij}\right)U_{Zj}
\end{align}

\makebox[\linewidth]{\rule{\textwidth}{1pt}}
$$\b{ZZh}$$
$$(+\,0\,0)$$
\begin{align}
     \vcenter{\hbox{\includegraphics[scale=1]{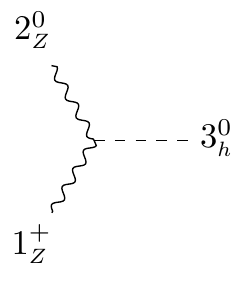}}} &\equiv  U_{Zi}\,U_{Hj}\,\mathcal{O}_{Z3} \vcenter{\hbox{\includegraphics[scale=1]{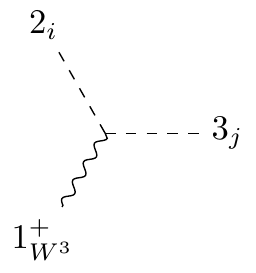}}} +   U_{Zi}\,U_{Hj}\,\mathcal{O}_{ZB}\vcenter{\hbox{\includegraphics[scale=1]{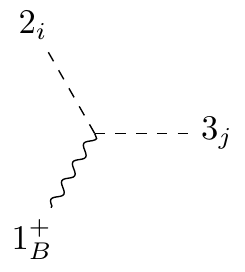}}}\nonumber
\end{align}
\begin{align}
   \f{e_{\scaleto{ZZH}{4pt}}}{2} \frac{\la12\ra\la31\ra}{\la23\ra} &\equiv  U_{Zi} \left(g\, \mathcal{O}_{Z3} \, T^3_{ij} \,+\, g^\prime\ \mathcal{O}_{ZB} \, T^B_{ij} \, \right) U_{hj}\frac{\la12\ra\la31\ra}{\la23\ra} \nonumber \\
   \implies \f{e_{\scaleto{ZZH}{4pt}}}{2} &= U_{Zi} \left( g\,\text{cos }\theta_w \, T^3_{ij} - g^\prime \, \text{sin }\theta_w \, T^B_{ij}\, \right) U_{Hj}
\end{align}

\makebox[\linewidth]{\rule{\textwidth}{1pt}}
$$\b{W^+W^-h}$$
$$(+\,0\,0)$$
\begin{align}
     \vcenter{\hbox{\includegraphics[scale=1]{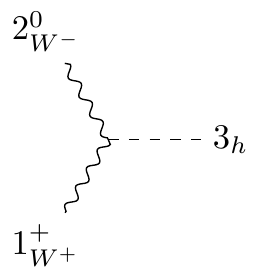}}} &\equiv U_{W^-i} \, U_{hj} \vcenter{\hbox{\includegraphics[scale=1]{figs/WpPhiPhi.pdf}}} \nonumber
\end{align}
\begin{align}
    \f{e_{\scaleto{WWH}{4pt}}}{2} \frac{\la12\ra\la31\ra}{\la23\ra} &\equiv g\, U_{W^-i} \, T^+_{ij} \, U_{hj} \frac{\la12\ra\la31\ra}{\la23\ra} \nonumber\\
    \implies   \f{e_{\scaleto{WWH}{4pt}}}{2} &= g\, U_{W^-i}T^+_{ij} \, U_{hj}
\end{align}
\makebox[\linewidth]{\rule{\textwidth}{1pt}}

\ndt This set of equations can be solved  by the ansatz
\bea
\label{eq:ansatz}
U_{W^+} = \frac{g}{\sqrt{2} m_{\scaleto{W}{4pt}}} \,T^+\cdot V \qquad U_{W^-} = \frac{g}{\sqrt{2}m_{\scaleto{W}{4pt}}} \, T^-\cdot V \nonumber\\
U_{Z} = \frac{1}{\sqrt{2}m_{\scaleto{Z}{4pt}}}\left(g\, \text{cos }\theta_w\, T^3_{ij} - g'\, \text{sin }\theta_w \, T^B_{ij}\right) \cdot V 
\eea
where $V = \lbrace v_1, v_2, v_3, v_4 \rbrace$. Note that despite the similarity of this equation with the usual Lagrangian based description of the Higgs mechanism, $V$ does not have the interpretation as the vacuum expectation value of scalar field here. The solution is
\bea
\label{eq:mathcingsol}
&& \hspace{1.5cm} v_1 =0 \qquad v_2 = 0 \qquad g' = g\, \text{tan }\theta_w \nn\\ 
&& m_{\scaleto{Z}{4pt}} = \frac{g}{2\cos \theta_w}\sqrt{v_3^2+v_4^2} \qquad m_{\scaleto{W}{4pt}} = \frac{g}{2}\sqrt{v_3^2+v_4^2} 
\eea
We get the exact solutions as the Standard Model because we have restricted the form of the three point amplitude in eq.(\ref{eq:WWZ}). Allowing for other structures will generalize the relation between $m_{\scaleto{Z}{4pt}}$ and $m_{\scaleto{W}{4pt}}$. Further note that when $g' \rightarrow 0$, we have $\theta_w = 0$ and $m_{\scaleto{Z}{4pt}} = m_{\scaleto{W}{4pt}}$. Here, we see the emergence of the custodial $SU(2) = SO(3)$. The three particles $W^\pm, Z$ all have equal mass in the limit where the hypercharge coupling vanishes.
\section{Four point amplitudes in the Electroweak sector}\label{sec:4ptamplitudes}
As we explained in the previous section, the structure of three point amplitudes is is severely restricted by Poincare' invariance and little group constraints. The construction of four point amplitudes from the three point ones requires more work.  Translation invariance is assured by the delta function in eq.(\ref{eq:momconservation}) and Lorentz invariance is guaranteed if we build the amplitude from the invariants in eq.(\ref{eq:masslessinv}) and eq.(\ref{eq:massiveinv}). These amplitudes must be little group tensors of the appropriate rank (or in the case of massless particles have appropriate little group weights). This still leaves open a multitude of possibilities. But beyond three points, we have new constraints arising from unitarity. The amplitude must factorize consistently on all the poles, i.e. when some subset of the external momenta goes on shell, the residue on the corresponding pole must factorize into the product of appropriate lower point amplitudes. In particular, if the exchanged particle is massless, we must have
\begin{align}
M \rightarrow \frac{M_L^{a \: h} M_R^{a\: -h}}{P^2}.
\end{align}
Here and below, $a$ is an index for the intermediate particle. In cases where there are particles which may have identical helicity and mass, this index distinguishes between them. Similarly for the exchange of a particle with mass $m$ and spin $S$, we have
\begin{align}
    M &\rightarrow \frac{M_L^{a\{I_1,\dots,I_{2S}\}} M_{R\{I_1,\dots,I_{2S}\}}^{a}}{P^2 - m^2} = \frac{ M_{L\{I_1,\dots,I_{2S}\}}^{a}\epsilon^{I_1 J_1}\dots \epsilon^{I_{2S}J_{2S}}  M_{R\{J_1,\dots,J_{2S}\}}^{a}}{P^2 - m^2}
\end{align}
For the rest of this section, we will work with four particle amplitudes with particles 1 and 2 incoming and 3 and 4 outgoing. Diagrammatically,
\begin{align}
 &\vcenter{\hbox{\includegraphics[scale = 1]{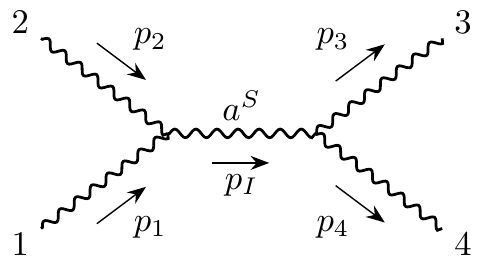}}} &= &\vcenter{\hbox{\includegraphics[scale = 1]{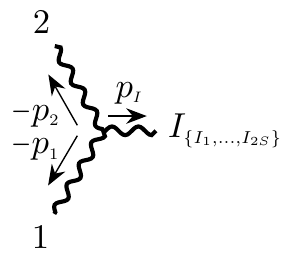}}} &\times& \frac{\epsilon^{I_1 J_1}\dots \epsilon^{I_{2S}J_{2S}}  }{p_I^2 - m^2} 
  &\times&\vcenter{\hbox{\includegraphics[scale = 1]{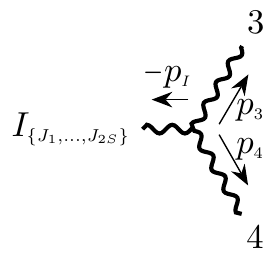}}} 
\end{align}
At four points, there are only three possible factorization channels defined by
\bea
\label{eq:factorization channels}
s = (p_1+p_2)^2 \qquad u = (p_1 - p_3)^2 \qquad t = (p_1-p_4)^2
\eea
We must ensure that the four point amplitude factorizes into appropriate three point amplitudes on all these channels. We do this by computing the residues in the s, t and u channels and 
\beas
\left( \frac{R_s}{s - m_s^2}+\frac{R_t}{t - m_t^2}+\frac{R_u}{u- m_u^2}\right)
\eeas 
where $m_s, m_t, m_u$ are the masses of the particles exchanged in the $s, t, u$ channels respectively. This procedure will yield local amplitudes for almost all cases. Only in the case of the $W^+W^-\gamma$ amplitude, which has one massless particle and two particles of equal mass, this yields a four point amplitude with $x$ factors which must be eliminated to get a local expression. We will go into more details in the corresponding section.\\

\ndt This represents only the factorizable part of the four point amplitude. We will find that these need to be supplemented by contact terms which depend on the specific form of the three point vertices. We can determine these by specifying the UV behaviour of the four point amplitudes. For the case of the Standard Model, we demand that they do not have any terms which grow with energy. This lets us determine the required contact terms. The complete four point amplitude is then written as 
\beas
M_4 = \left( \frac{R_s}{s - m_s^2}+\frac{R_t}{t - m_t^2}+\frac{R_u}{u- m_u^2}\right) + P(\lambda_i, \tilde{\lambda}_i)
\eeas
where $P$ is a Lorentz invariant polynomial in the spin spinors corresponding to the four particles with the appropriate number of little group indices.
\subsection{\texorpdfstring{$W^+ W^- \rightarrow W^+ W^-$}{WW to WW}} \label{sec:wwscattering}
In this section, we analyze the scattering of $W^+ W^-\rightarrow W^+W^-$. For the sake of explicit calculations, we make the following choice for the 4 particle kinematics (with particles 1, 2, 3 and 4 corresponding to $W^-, W^+, W^+, W^-$ respectively. 
\begin{alignat}{2}
    \label{eq:WWWWkinematics}
    p_1 &= (E, 0, 0, p) &\quad\quad p_2 &= (E, 0, 0, -p) \\
     p_3 &= (E, p \sin\theta, 0, p  \cos \theta) &\quad\quad p_4 &= (E, -p \cos \theta, 0, -p \cos\theta) \nonumber
\end{alignat}

We will see that, based on the three point amplitudes listed in Section~[\ref{sec:IR}], the scattering can occur in the $s$ and $t$ channels via the exchange the $Z$, $A$ or $h$.
\begin{align}
    \vcenter{\hbox{\includegraphics[scale=.65]{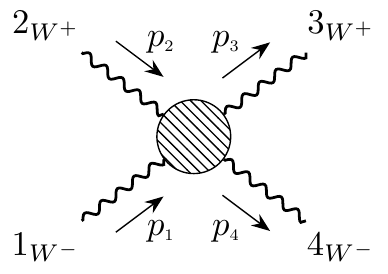}}} &\supset\vcenter{\hbox{\includegraphics[scale=.65]{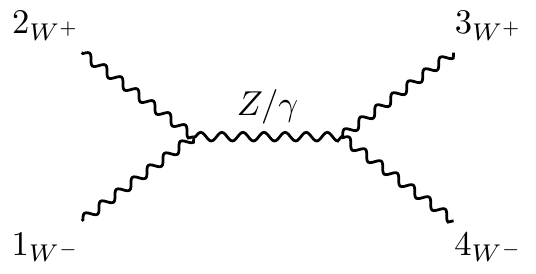}}} + \vcenter{\hbox{\includegraphics[scale=.65]{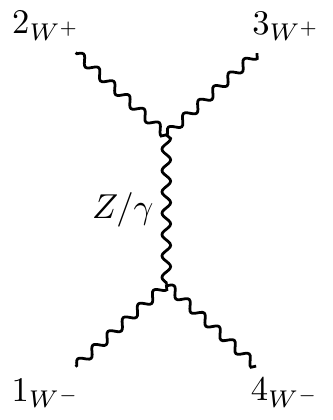}}} + \vcenter{\hbox{\includegraphics[scale=.65]{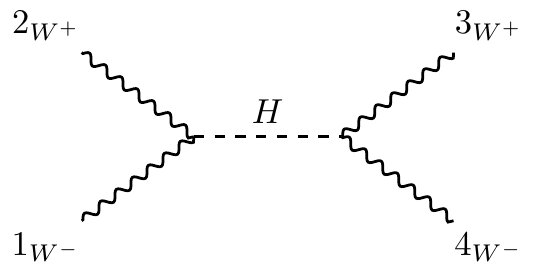}}} + \vcenter{\hbox{\includegraphics[scale=.65]{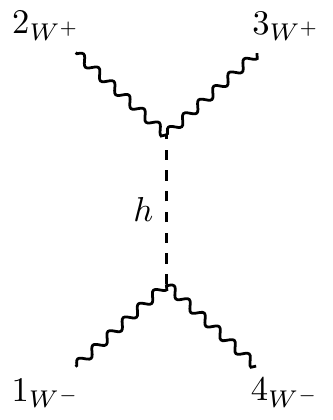}}}  
\end{align}

\subsubsection{\texorpdfstring{$s$ - Channel}{s-Channel}} 
\label{sec:schannel}
\begin{itemize}
\item Z exchange
\end{itemize}
We can glue together two $W^+W^-Z$ three point amplitudes and construct the residue in the $s$ - channel. 
\begin{align}
  ( M^Z_{L})^{\lbrace I_1I_2\rbrace} &= \frac{e_{\scaleto{W}{4pt}}}{m_{\scaleto{W}{4pt}}^2 m_{\scaleto{Z}{4pt}}}\left(\la\b{12}\ra[\b{12}]\la\b{I}|(-p_1) - (-p_2)|\b{I}] + \text{cyc.} \right)^{\lbrace I_1I_2\rbrace} \qquad \\
   ( M^Z_{R} )_{\lbrace I_1I_2\rbrace}  &= \frac{e_{\scaleto{W}{4pt}}}{m_{\scaleto{W}{4pt}}^2 m_{\scaleto{Z}{4pt}}}\left(- \la\b{34}\ra[\b{34}]\la\b{I}|p_3 - p_4|\b{I}] + \text{cyc.} \right)_{\lbrace I_1I_2\rbrace} 
\end{align}
Here $I = p_1+p_2$ is the momentum exchanged and we have suppressed the little group indices corresponding to the external particles. The residue on the $s$ - channel is  
\beas
R_s^Z =  (M^Z_L)^{\lbrace I_1I_2\rbrace} (M_R^Z)_{\lbrace I_1I_2\rbrace}
\eeas
Evaluating this expression yields
\begin{align}
\label{eq:zresidue}
   R_s^Z &= \frac{e_{\scaleto{W}{4pt}}^2}{m_{\scaleto{W}{4pt}}^4}\Bigg\lbrace \, 2 \,\la\b{12}\ra \, [\b{12}] \, \la\b{34}\ra \, [\b{34}]\, (p_1 - p_2).(p_3 - p_4) \\
   &\hspace{1.4cm}+ 4 \Big(\la \b{42}\ra \,  [\b{24}] \,\la\b{1}|p_2|\b{1}] \,\la\b{3}|p_4|\b{3}] + \la \b{31}\ra \, [\b{13}] \,\la\b{2}|p_1|\b{2}] \, \la\b{4}|p_3|\b{4}] - (1\, \leftrightarrow \,2)\Big)\nonumber \\
      &\hspace{1.4cm}+ 2 \,\Big( \la \b{12} \,\ra[\b{12}]\, \la\b{4}|p_3|\b{4}]\, \la \b{3}| p_1-p_2 |\b{3}] \, + \, \la \b{12}\ra \, [\b{12}]\, \la \b 3 |p_4|\b 3] \,\la \b 4|p_1-p_2|\b 4]  -(1\, \leftrightarrow \, 2) \Big)\Bigg\rbrace\nn 
\end{align}
The full details of the calculation are presented in Appendix [\ref{sec:gluing}]. 
\begin{itemize}
\item Photon exchange
\end{itemize}
This corresponds to gluing together the two $W^+W^-\gamma$ vertices. There are two possibilities
\begin{align}
\label{eq:photonexchange}
  M_L^{-} &= \frac{e}{m_{\scaleto{W}{4pt}}} x_{12}^-\,\, \b{\la12\ra}^2 \qquad M_R^{+} = \frac{e}{m_{\scaleto{W}{4pt}}} x_{34}^+\,\, \b{[34]}^2 \\ \nonumber
  M_L^{+} &= \frac{e}{m_{\scaleto{W}{4pt}}} x_{12}^+\,\, \b{[12]}^2 \qquad M_R^{-} = \frac{e}{m_{\scaleto{W}{4pt}}} x_{34}^-\,\, \b{\la 34\ra}^2 
\end{align}
where the superscripts indicate the helicity of the photon. Note that the definition of $x$-factors differs slightly from Appendix [\ref{sec:mincoupling}] due to the fact that $p_1$ and $p_2$ are now incoming momenta. The appropriate definitions are 
\begin{align*}
   \frac{(-p_1+p_2)_{\alpha\dot\alpha}}{2m} \lambda_I^\alpha &= x_{12}^+\,\, \tilde\lambda_{I\dot\alpha} &\quad  \frac{(p_3-p_4)_{\alpha\dot\alpha}}{2m} \lambda_I^\alpha &= x_{34}^+ \,\,(-\tilde\lambda_{I\dot\alpha}) \\
   \frac{(-p_1+p_2)_{\alpha\dot\alpha}}{2m} \tilde\lambda_I^{\dot\alpha} &= x_{12}^-\,\, \lambda_{I\alpha} &\quad   \frac{(p_3-p_4)_{\alpha\dot\alpha}}{2m} (-\tilde\lambda_I^{\dot\alpha})   &= x_{34}^-\,\, \lambda_{I\alpha}
\end{align*}
The extra minus sign that accompanies $\tilde{\lambda}_{I\dot{\alpha}}$ in the equations defining $x_{34}^\pm$ is because the momentum $I$ is incoming. The residue corresponding to the photon exchange is a sum over both the possibilities in eq.(\ref{eq:photonexchange}).
\begin{align}
   R_s^A &=\frac{e^2}{m_{\scaleto{W}{4pt}}^2} \left(x_{12}^- x_{34}^+ \la\b{12}\ra^2[\b{34}]^2 + x_{12}^+x_{34}^-[\b{12}]^2\la\b{34}\ra^2\right)
\end{align}
We must now eliminate the $x$ - factors in order to obtain a local expression for this residue. There are multiple ways to achieve this and they generally result in different expressions for the residue. It is important to emphasize that while these forms are precisely equal on the factorization channel, they all lead to different expressions away from the pole. Since the physical amplitude must be the same, they yield different contact terms. The complete details of the calculation are delegated to Appendix~[\ref{sec:gluing}]. Here, we present two different expressions for the residue on the $s - $channel.
\begin{align}\label{eq:photonexchange}
R_s^{\gamma} &= \frac{e^2}{2m_{\scaleto{W}{4pt}}^4}\Bigg\lbrace  (p_1-p_2).(p_3-p_4) \, \b{\la12\ra \, [12] \, \la 34 \ra \, [34]\,} \, \\
&\hspace{1.6cm} +\, \Big(\b{\la 12 \ra [12]\Big[ \la 3|}\, (p_1 - p_2)\, (p_1 + p_2)\,|\b{4\ra \, [34]} - \la \b{34\ra \,  [3|}\,(p_1 + p_2) \, (p_1 - p_2) \, \b{|4]}\Big] \nn \\
& \hspace{1.6cm}- \b{\la 1|} \, p_1+p_2 \, | \b{4] \, [3|} \, p_1 + p_2 \, |\b{2\ra \, [12] \, \la34\ra} + (1, \, 2 \leftrightarrow 3, 4 )\Big) \Bigg\rbrace \nn
\end{align}
This expression can be manipulated to look identical to eq.(\ref{eq:zresidue}). This requires the use of the following Sch$\ddot{\text{o}}$uten identities
\beas
&&\la \b {3|} \, (p_1 - p_2) \, I \, |\b{4\ra \, [34]}  - \la \b{34\ra \, [3|}\,I \, (p_1 - p_2) \, \b{|4]} \\
&& =2 \Big(\la \b 4| \, p_3 \, | \b 4]  \, \la\b 3 |\, (p_1 - p_2)\, | \b 3]\, -  ( 3 \leftrightarrow 4) \Big)
\eeas
and
\beas
&& \b{\la 1|}\, I \, | \b{4] \, [3|} \,I \,|\b{2\ra \, [12] \, \la 34 \ra} + (1, \, 2 \leftrightarrow 3, 4 ) \\
&&= 4 \Big(\la \b{42}\ra \, [\b{24}] \, \la\b{1}| \, p_2 \, |\b{1}] \, \la\b{3}| \, p_4 \, |\b{3}]\,  +\,  \la \b{31}\ra \,  [\b{13}]  \, \la\b{2}|\, p_1 \,|\b{2}] \, \la\b{4}|\, p_3 \, |\b{4}] \, -\, (1\, \leftrightarrow \, 2)\Big)
\eeas
where $I = p_1 + p_2$. These identities are true only on the factorization channel $I^2 = 0$ on which we can write $R_s^\gamma = R_s^Z \, (m_Z = 0)$. This is not true away from the factorization channel. Consequently the contact terms that must be added to achieve the correct UV behaviour differ. This explicitly demonstrates the dependence of contact terms on the specific form of the three point amplitudes.
\begin{itemize}
\item Higgs exchange
\end{itemize}
This is the simplest to compute. We just glue together the following amplitudes.
\begin{align}
  M_L &= \frac{e_{\scaleto{WWH}{4pt}}}{m_{\scaleto{W}{4pt}}} \b{\la12\ra[12]} \qquad M_R = \frac{e_{\scaleto{WWH}{4pt}}}{m_{\scaleto{W}{4pt}}} \b{\la34\ra[34]}
\end{align}
which directly yields 
\begin{align}
  R_s^h &= \frac{e_{\scaleto{WWH}{4pt}}^2}{m_{\scaleto{W}{4pt}}^2}\b{\la12\ra[12]\la34\ra[34]}
\end{align}
The complete contribution of the $s - $ channel is
\beas
M_s = \left( \frac{R_s^Z}{s - m_{\scaleto{Z}{4pt}}^2} +  \frac{R_s^{\gamma}}{s} +  \frac{R_s^h}{s - m_h^2} \right)
\eeas
\subsubsection{\texorpdfstring{$t$ - channel}{t - channel}}
The computation of the $t$ - channel residues is very similar to that of the $s$ - channel. In fact, we can obtain them from the $s - $ channel ones by the replacement $p_2 \leftrightarrow -p_4$. The results are presented below with $I = p_1 - p_4$.
\begin{itemize}
\item Z exchange
\end{itemize}
\begin{align}
\label{eq:ztresidue}
   R_t^Z &= \frac{e_{\scaleto{W}{4pt}}^2}{m_{\scaleto{W}{4pt}}^4} \, \Bigg\lbrace 2 \, \la\b{14}\ra \, [\b{14}] \, \la\b{32}\ra \, [\b{32}] \, (p_1 + p_4).(p_3 + p_2)  \, \\
   &\hspace{1.4cm} + 4 \Big(\la \b{31}\ra \, [\b{13}] \, \la\b{4}| \, p_1 \, |\b{4}] \, \la\b{2}| \, p_3 \, |\b{2}] - \la \b{42}\ra \, [\b{24}] \, \la\b{1}| \, p_4 \, |\b{1}] \, \la\b{3}| \, p_2 \, |\b{3}] - (1\, \leftrightarrow \,4)\Big)\nonumber \\
      &\hspace{1.4cm} + 2\Big(\la \b{12}\, \ra[\b{14}]\, \la\b{2}| \,p_3 \,|\b{2}] \, \la \b{3}|\, p_1 + p_2\, |\b{3}] + \la \b{14}\ra \, [\b{14}]\, \la \b 3 |\, p_2 \,|\b 3] \, \la \b 2| \, p_1 + p_4 \,|\b 2] - (1\, \leftrightarrow \, 4) \Big)\nn  \Bigg\rbrace
\end{align}
\begin{itemize}
\item Photon exchange
\end{itemize}
The residue on the $t - $ channel resulting from gluing together two $W^+W^-\gamma$ amplitudes is
\begin{align}\label{eq:tphotonexchange}
R^\gamma_t &= \frac{e^2}{2m_{\scaleto{W}{4pt}}^4} \,\Bigg\lbrace  -(p_1 + p_4).(p_3 + p_2) \,\b{\la 14 \ra \, [14] \, \la 32 \ra \ [32]} \\ 
&\hspace{1.6cm}+\Big(\b{\la 14 \ra\, [41]\,\Big[ \la 3|}\, (p_1 + p_4) \, (p_1 - p_4) \,\b{2\ra \, [32]}  - \la \b{32\ra \,  [3|}\,(p_1 - p_4) \, (p_1 + p_4)\,\b{|2]} \,\Big] \nn \\ 
&\hspace{1.6cm} + \b{\la 1|}\, p_1 - p_4 \,| \b{2] \, [3|}\, p_1 - p_4 \,|\b{4\ra \, [14] \, \la 32 \ra} + (1, \, 4 \leftrightarrow 3, 2 )\Big) \Bigg\rbrace  \nn
\end{align}
\begin{itemize}
\item Higgs exchange
\end{itemize}
\begin{align}
  R_t^h &= \frac{e_{\scaleto{WWH}{4pt}}^2}{m_{\scaleto{W}{4pt}}^2}\b{\la 14 \ra \, [14] \,\la 23 \ra \, [23]}
\end{align}
The total contribution from the $t - $channel is
\beas
M_t = \left( \frac{R_t^Z}{t - m_{\scaleto{Z}{4pt}}^2} +  \frac{R_t^{\gamma}}{t} +  \frac{R_t^h}{t - m_h^2} \right)
\eeas
\subsubsection{Contact terms}\label{sec:WWcontactterms}
The quantity $M \equiv M_s+M_t $ has been constructed to have the correct factorization properties. As explained before, the behaviour away from the factorization channels depends on the specific forms of the three point amplitudes. We can impose further constraints on the amplitude to fix it completely. It is evident that the high energy limit of the amplitude is ill defined due to the presence of the $\frac{1}{m^4_W}$ poles which leads to amplitudes which grow with energy as $E^4$. This violates perturbative unitarity. If we insist that the theory has a well defined high energy limit, we must add contact terms (which by definition have 0 residue on the factorization poles) to cancel this $E^4$ growth. The form of the contact terms can be deduced by figuring out which components of the amplitude grow in the UV. Plugging in the 4-particle kinematics in $\ref{eq:WWWWkinematics}$, we find that only the all longitudinal component grows as $E^4$, 
\begin{align}
 M &\rightarrow \frac{E^4}{8m_{\scaleto{W}{4pt}}^4} \left(e^2+e_{\scaleto{W}{4pt}}^2\right)\left(-5-12 \text{ cos }\theta + \text{ cos }2\theta \right) \, .
\end{align} 
The following contact term serves to kill these high energy growths
\bea
\label{eq:wwcontact}
\hspace{-6mm}c_{\scaleto{WWWW}{4pt}}=&& \frac{e^2+e_{\scaleto{W}{4pt}}^2}{m_{\scaleto{W}{4pt}}^4}\left( -\la \b{12}\ra [\b{12}]\la \b{34}\ra [\b{34}] + 2 \la \b{13}\ra [\b{13}]\la \b{24}\ra [\b{24}] - \la \b{14}\ra [\b{14}]\la \b{23}\ra [\b{23}]  \right)
\eea
Adding these contact terms, we find that the amplitude still grows as $E^2/m_{\scaleto{W}{4pt}}^2$. Demanding that the coefficient of this growing term vanishes enforces $e_{\scaleto{WWH}{4pt}}^2 = e^2 + e_{\scaleto{W}{4pt}}^2$.
\subsection{\texorpdfstring{$ W^+  Z \rightarrow  W^+ Z$}{WZ to WZ}}\label{sec:WZ scattering}
The 4 particle kinematics appropriate to this situation is 
\bea
\label{eq:WZWZkinematics}
&& p_1 = (E_1, 0, 0, p) \hspace{3.4cm} p_2 = (E_2, 0, 0, -p) \\
&&\nonumber p_3 = (E_2, p \text{ sin }\theta, 0, p \text{ cos }\theta) \qquad\qquad p_4 = (E_1, -p \text{ sin }\theta, 0, -p \text{ cos }\theta)
\eea
This configuration automatically satisfies momentum conservation. We can rewrite $E_2$ in terms of $E_1$ by using the on-shell constraint as $E_2 = \sqrt{E_1^2-m_{\scaleto{W}{4pt}}^2+m_{\scaleto{Z}{4pt}}^2}$. We can build this amplitude by gluing together two $W^+W^-Z$ amplitudes in two ways and by gluing two $W^+W^-h$ amplitudes.
\begin{align}
    \vcenter{\hbox{\includegraphics[scale=.8]{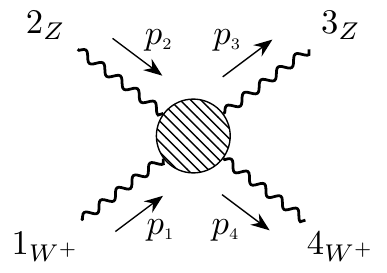}}} &\supset\vcenter{\hbox{\includegraphics[scale=.8]{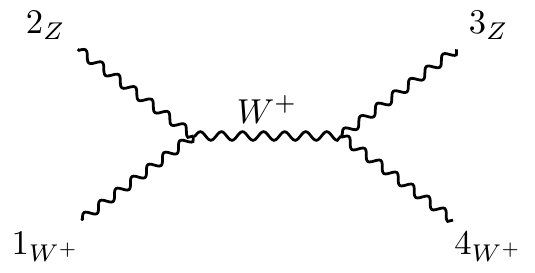}}} + \vcenter{\hbox{\includegraphics[scale=.8]{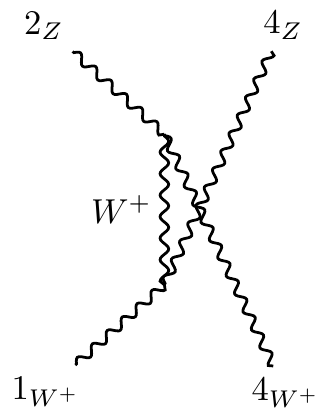}}} + \vcenter{\hbox{\includegraphics[scale=.8]{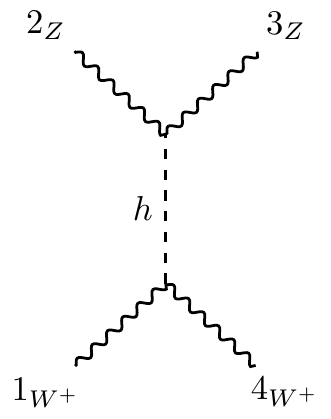}}} \label{eq:WZtoWZ}  \
\end{align}
We present the final expressions below. The calculations are very similar to those involved in $W^+W^-\rightarrow W^+W^-$.
\begin{itemize}
\item $s - $channel W - exchange
\end{itemize}
\begin{align}\label{eq:wzsexchange}
 R_s^W&= \frac{e_{\scaleto{W}{4pt}}}{m_{\scaleto{W}{4pt}}^4m_{\scaleto{Z}{4pt}}^2} \Bigg\lbrace  2 (m_{\scaleto{W}{4pt}}^2 - m_{\scaleto{Z}{4pt}}^2)^2 \b{\la 12 \, \ra [12] \,  \la 34 \ra \,  [34] \, }  m_{\scaleto{W}{4pt}}^2 \big( 2\b{\la 12 \ra \,  [12] \,  \la 34 \ra \, [34] \, }(p_1 - p_2).(p_3-p_4)  \nn\\  
   & \hspace{1.6cm}+4\Big[ \la \b{42}\ra \,  [\b{24}] \, \la\b{1}|p_2|\b{1}] \, \la\b{3}|p_4|\b{3}] - \la \b{32}\, \ra[\b{23}]\, \la\b{1}|p_2|\b{1}]\, \la\b{4}|p_3|\b{4}]+ (1, 3 \leftrightarrow 2, 4 )\Big]\nn \\
    &\hspace{16mm}+ 2\Big[ \la \b{12} \ra[\b{12}] \Big(\la\b{4}|p_3|\b{4}]\, \la \b{3}| p_1-p_2 |\b{3}] \, - (3 \leftrightarrow 4)\Big)+ (1, 2 \leftrightarrow 3, 4) \Big]\Bigg\rbrace
\end{align}
\begin{itemize}
\item $u - $channel W - exchange
\end{itemize}
\begin{align}\label{eq:wzuexchange}
 R_u^W&= \frac{e_{\scaleto{W}{4pt}}}{m_{\scaleto{W}{4pt}}^4m_{\scaleto{Z}{4pt}}^2} \Bigg\lbrace  2 (m_{\scaleto{W}{4pt}}^2 - m_{\scaleto{Z}{4pt}}^2)^2 \b{\la 13 \, \ra [13] \,  \la 24 \ra \,  [24] \, }  m_{\scaleto{W}{4pt}}^2 \big(- 2\b{\la 13 \ra \,  [13] \,  \la 24 \ra \, [24] \, }(p_1 + p_3).(p_2 + p_4)  \nn\\  
   & \hspace{1.6cm}+4\Big[-\la \b{43}\ra \,  [\b{34}] \, \la\b{1}|p_3|\b{1}] \, \la\b{2}|p_4|\b{2}] - \la \b{32}\, \ra[\b{23}]\, \la\b{1}|p_3|\b{1}]\, \la\b{4}|p_2|\b{4}]+ (1, 2 \leftrightarrow 3, 4 )\Big]\nn \\
    &\hspace{16mm}+ 2\Big[- \la \b{13} \ra[\b{13}] \Big(\la\b{4}|p_2|\b{4}]\, \la \b{2}| p_1+p_3 |\b{3}] \, + (2 \leftrightarrow 4)\Big)+ (1, 3 \leftrightarrow 2, 4) \Big]\Bigg\rbrace
\end{align}
\begin{itemize}
\item $t - $channel Higgs - exchange
\end{itemize}
\begin{align}
R_t^h &= \frac{e_{\scaleto{WWH}{4pt}}}{m_{\scaleto{W}{4pt}}}\frac{e_{\scaleto{ZZH}{4pt}}}{m_{\scaleto{Z}{4pt}}} \b{\la23\,\ra[23]\,\la14\,\ra[14]} \, .
\end{align}
\begin{itemize}
\item Contact terms
\end{itemize}
We are again in the familiar situation where the quantity 
\beas
\left(\frac{R_s^W}{s - m_{\scaleto{W}{4pt}}^2} + \frac{R_u^W}{u - m_{\scaleto{W}{4pt}}^2} + \frac{R_t^h}{t - m_h^2}\right) \, ,
\eeas
factorizes correctly on all the factorization channels. However, the all longitudinal component again grows with energy as can be seen by evaluating this using the kinematics in eq.(\ref{eq:WZWZkinematics}). We find that the following contact term is needed to fix this and have a well behaved theory in the UV,
\begin{align}
c_{\scaleto{WZWZ}{4pt}} = \frac{e_{\scaleto{W}{4pt}}^2}{m_{\scaleto{W}{4pt}}^4} (\b{\la 12 \ra\, [12]\, \la 34\ra\, [34] + \la 23\ra \, [23] \, \la 14\ra \, [14] - 2 \la 24\ra\, [24]\,\la13\ra\,[13]}) \,.
\end{align}
Furthermore, to kill growth at $\mo (E^2)$, we must also have $e_{\scaleto{WWH}{4pt}}\, e_{\scaleto{ZZH}{4pt}} \, m_{\scaleto{W}{4pt}}^3 = e_{\scaleto{W}{4pt}}^2 \, m_{\scaleto{Z}{4pt}}^3$. 

\subsection{\texorpdfstring{$W^+W^- \rightarrow Zh$}{WW to Zh}}
We next consider the scattering $W^+W^- \rightarrow Zh$ with the following kinematics
\bea
\label{eq:WWZhkinematics}
&&  p_1^\mu = (E_1, 0, 0, p_1) \hspace{3.4cm} p_2^\mu = (E_1, 0, 0, -p_1) \\
&&\nonumber p_3^\mu = (E_3, p_2 \text{ sin }\theta, 0, p_2 \text{ cos }\theta) \qquad\qquad p_4^\mu = (E_4, -p_2 \text{ sin }\theta, 0, -p_2 \text{ cos }\theta)
\eea
Using on-shell constraints, we can eliminate $p_1, p_2, E_3, E_4$ in favor of $E_1$.
\beas
p_1 = \sqrt{E_1^2-mW^2} \quad p_2 = \sqrt{E_2^2-m_{\scaleto{Z}{4pt}}^2} \quad E_3 = \frac{m_{\scaleto{Z}{4pt}}^2 - m_h^2 + 4 E_1^2}{4 E_1}, E_4 = \frac{4 E_1^2 - m_{\scaleto{Z}{4pt}}^2 + m_h^2}{4 E_1}
\eeas
We can build this amplitude by gluing together $(W^+W^-Z, Zhh)$ on the $s - $channel and by gluing together $(W^+W^-h, ZZh)$ in the $u$ and $t$ channels as shown
\begin{align}
    \vcenter{\hbox{\includegraphics[scale=.8]{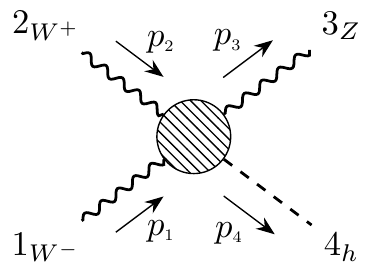}}} &\supset \vcenter{\hbox{\includegraphics[scale=.8]{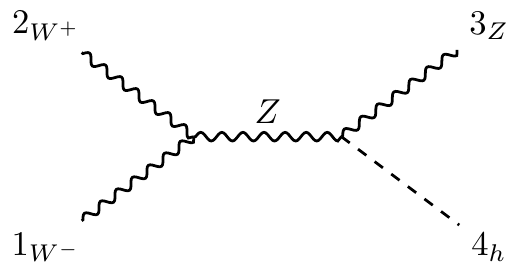}}} + \vcenter{\hbox{\includegraphics[scale=.8]{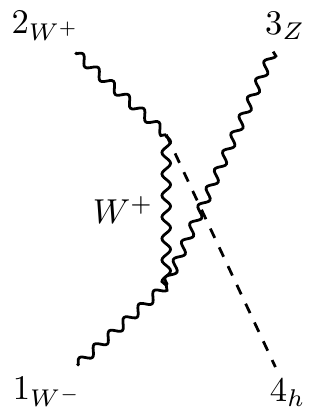}}} + \vcenter{\hbox{\includegraphics[scale=.8]{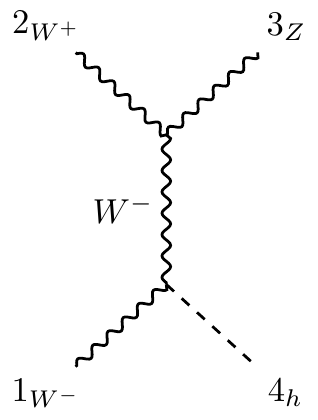}}}   \
\end{align}
Using the familiar procedure, we get
\begin{itemize}
\item $s - $ channel Z exchange
\end{itemize}
\begin{align}
R^Z_s &= \frac{e_{\scaleto{ZZH}{4pt}}\, e_{\scaleto{W}{4pt}}}{m_{\scaleto{W}{4pt}}^2} \left( -\la\b{12}\ra\, [\b{ 12}]\, \la\b{3}|p_1-p_2|\b{3}] - 2 \la \b{23}\ra \, [\b{23}]\, \la\b{1}|p_2|\b{1}] + 2 \la\b{13}\ra\,[\b{13}]\la\,\b{2}|p_1|\b{2}] \right)
\end{align}
\begin{itemize}
\item  $u - $ channel $W$ exchange
\end{itemize}
\begin{align}
R_u^W &= \frac{e_{\scaleto{W}{4pt}} \,\, e_{\scaleto{WWH}{4pt}}}{m_{\scaleto{W}{4pt}} m_{\scaleto{Z}{4pt}}}\left(-\b{\la 13 \ra \, [13] \la 2 |} p_1 + p_3 \b{| 2]} + 2 \b{\la 23 \ra \,[ 23 ] \la 1|}p_3\b{|1]} - 2 \b{ \la 12 \ra \, [ 12] \la 3}|p_1|\b{3]} \right) 
\end{align}
\begin{itemize}
\item $t - $ channel $W$ exchange
\end{itemize}
\begin{align}
R_t^W &=  \frac{e_{\scaleto{W}{4pt}}\,\, e_{\scaleto{WWH}{4pt}}}{m_{\scaleto{W}{4pt}} m_{\scaleto{Z}{4pt}}} \left( - \b{\la 23 \ra \, [23] \, \la 1 |}p_2 + p_3 \b{|1]} + 2 \b{ \la13\ra \, [13] \la2 | }p_3 \b{| 2 ] } - 2 \b{\la 12\ra \, [12] \, \la 3|}p_2\b{|3]} \right)
\end{align}
\begin{itemize}
\item Contact terms
\end{itemize}
In this case, the component of the amplitude with $W^+, W^-, Z$ all being longitudinal grows with energy. However, there are no possible contact terms that are compatible with Loretnz invriance and the little group. The vanishing of the growing term imposes a constraint on the couplings $e_{\scaleto{WWH}{4pt}}, e_{\scaleto{ZZH}{4pt}}$.
\begin{align}
  \frac{e_{\scaleto{ZZH}{4pt}}}{e_{\scaleto{WWH}{4pt}}} = \frac{m_{\scaleto{W}{4pt}}}{m_{\scaleto{Z}{4pt}}}
\end{align}

\subsection{\texorpdfstring{$W^+W^- \rightarrow hh$}{WW to hh}}
To compute this amplitude, we can glue together two $W^+W^-h$ amplitudes in the $t$ and $u$ channels. 
\begin{equation}
    \vcenter{\hbox{\includegraphics[scale=0.8]{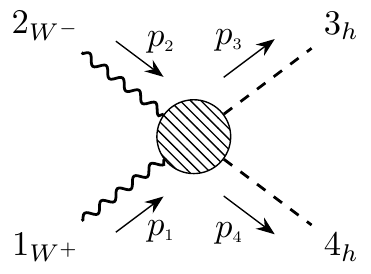}}} \supset \vcenter{\hbox{\includegraphics[scale=0.8]{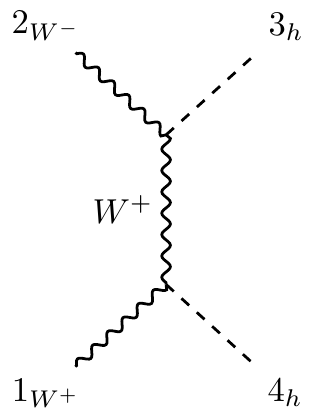}}} + \vcenter{\hbox{\includegraphics[scale=0.8]{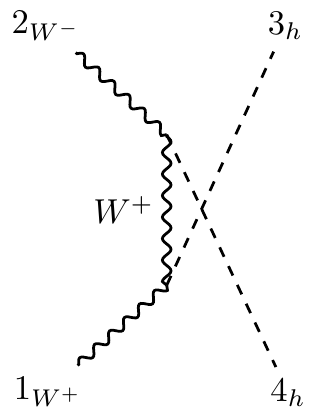}}}
\end{equation}
\begin{itemize}
\item $t$-channel $W$ exchange
\end{itemize}
\begin{equation}
R_t^W = \frac{e_{\scaleto{WWH}{4pt}}^2}{2m_{\scaleto{W}{4pt}}^2}\left(- 2m_{\scaleto{W}{4pt}}^2 \la \b {12}\ra \, [\b{12}]+\, m_{\scaleto{W}{4pt}} \, \la \b{12}\ra \la \b 1|p_4|\b 2] +m_{\scaleto{W}{4pt}} [\b{12}] [\b 1|p_4|\b 2\ra - \la \b 1|p_4|\b 2] [\b 1 |p_4|\b 2\ra\right)
\end{equation}
\begin{itemize}
\item $u - $ channel $W$ exchange
\end{itemize}
\begin{equation}
R_t^W = \frac{e_{\scaleto{WWH}{4pt}}^2}{2m_{\scaleto{W}{4pt}}^2}\left(- 2m_{\scaleto{W}{4pt}}^2 \la \b {12}\ra \, [\b{12}]+\, m_{\scaleto{W}{4pt}} \, \la \b{12}\ra \la \b 1|p_3|\b 2] +m_{\scaleto{W}{4pt}} [\b{12}] [\b 1|p_3|\b 2\ra - \la \b 1|p_3|\b 2] [\b1 |p_3|\b 2\ra\right)
\end{equation} 

\subsection{\texorpdfstring{$hh\rightarrow hh$}{hh to hh}}
The high energy limit of $W^+W^-\rightarrow W^+W^-$, yields a constant term. This shows that there is a 4 point amplitude for scattering of the scalars. Due to the $SO(4)$ symmetry of the scalar sector, we must also have a 4 point amplitude for the scattering of higgs particles. This can arise in the IR only if there is a $hhh$ coupling. This a trilinear higgs coupling is a necessary part of the theory.\\
\section{Conclusions and Outlook}
\ndt We have presented a completely on-shell description of the higgs mechanism within the Standard Model. We see that all the physics is reproduced by demanding consistent factorization, correct ultraviolet behaviour and consistency of the UV and IR. The precise relations between the masses of the $W^\pm, Z$ and $\theta_w$ depend on the structures that have been included in the three point $W^+W^-Z$ amplitude. Our choice of the three point amplitude in eq.(\ref{eq:WWZ}) ensured that we reproduced the usual result. We have constructed four particle, tree-level amplitudes from three particle amplitudes. The construction of higher point amplitudes and extensions to loop amplitudes are the obvious next questions. We have also restricted the particle content of the scalar sector to a single, real scalar transforming under an $SO(4)$ global symmetry. We have studied the Higgs mechanism for $SU(2)_L \times U(1)_Y$ breaking to $U(1)_{\text{EM}}$, relevant to electroweak symmetry breaking. It would be interesting to extend this analysis to completely general theories. \\

\ndt This work is a preliminary step in connecting modern methods in scattering amplitudes to the real world. There have been many developments in new ways of thinking about scattering amplitudes. It has proved useful to think of them as differential form on kinematic space \cite{diffforms}. These differential forms are associated to geometric structures in many cases. The physics of scattering amplitudes emerges from simple properties of the underlying geometry as seen in the few known cases \cite{polytope1, polytope2, polytope3, polytope4, polytope5}. It would therefore be useful to rewrite amplitudes in the Standard Model as differential forms. This would lay the groundwork for an attempt to look for hidden geometric structure within these amplitudes. \\
\acknowledgments
We thank Nima Arkani-Hamed for guidance at all stages of this project. We thank Wayne Zhao for useful discussions. BB is supported in part by the Government of the Republic of Trinidad and Tobago.
\appendix

\section{Conventions}
\label{sec:conventions}
In this appendix, we explicitly state all our conventions. We work with the metric signature $(+,-,-,-)$. $\mathrm{SU}(2)_L$ (of the bosons in the UV) and $\mathrm{SL}(2,\mathbf{C})$ spinor indices are raised and lowered using
\begin{equation}
    \epsilon_{\alpha\beta} = - \epsilon^{\alpha\beta}= 
    \begin{pmatrix}
    0 & -1 \\
    1 & 0
    \end{pmatrix}
\end{equation}
We raise and lower the $\mathrm{SL}(2,\mathbf{C})$ as follows
\beas
&&\lambda^\alpha = \epsilon^{\alpha \beta}\lambda_\beta  \qquad \lambda_\beta = \epsilon_{\beta \alpha }\lambda^\alpha \qquad \tilde{\lambda}^{\dot{\alpha}} = \epsilon^{\dot{\alpha} \dot{\beta}}\tilde{\lambda}_{\dot{\beta}} \qquad \tilde{\lambda}_{\dot{\beta}} = \epsilon_{\dot{\beta}\dot{\alpha}}\tilde{\lambda}^{\dot{\alpha}}\nn
\eeas
We use the same tensor for lowering and raising the indices of the massive little group $\mathrm{SU}_L(2)$ with the convention
\beas
\lambda_{\alpha I} = \lambda_\alpha^J \,\epsilon_{JI} \qquad \lambda_\alpha^I = \lambda_{\alpha J}\, \epsilon^{JI} .
\eeas  
Note that the Greek $\mathrm{SL}(2, \mathbf{C})$ spinor indices are raised and lowered on the left while the Latin little group indices are raised and lowered on the right. We also make use of angle and square spinors which are defined as 
\begin{align}\label{eq:massivesuppress}
    |\mathbf{i}\ra&:= |\mathbf{i}\ra_\alpha^I = \lambda_\alpha^I, &\qquad \la\mathbf{i}| &:= \la\mathbf{i}|^{\alpha I} = \lambda^{\alpha I} ,\\
    |\mathbf{i}] &:= |\mathbf{i}]^{\dot{\alpha}I} = \tilde{\lambda}^{\dot{\alpha}I}, &\qquad [\mathbf{i}| &:= [\mathbf{i}|_{\dot{\alpha}}^I = \tilde{\lambda}^I_{\dot{\alpha}}.\nn
\end{align}
With this the momentum can be written as,
\beas
    p_{\alpha\dot{\alpha}} = \epsilon_{JI}|\mathbf{i}\rangle^I [{\mathbf{i}}|^J = \epsilon_{JI}\lambda_\alpha^I \tilde{\lambda}_{\dot{\alpha}}^J, \qquad\qquad  p^{\dot{\alpha}\alpha} =\epsilon_{JI}\,\la\mathbf{i}|^{I}\, |\mathbf{i}]^J = \epsilon_{JI} \lambda^{\alpha I}\tilde{\lambda}^{\dot{\alpha}J}
\eeas
We follow the convention that undotted indices are contracted from top to bottom while dotted indices are contracted from the bottom to the top.
\begin{align}
    \la\mathbf{ij}\ra = \la\mathbf{i}|^{\alpha I} |\mathbf{j}\ra_\alpha^J, &\qquad [\mathbf{ij}] = [\mathbf{i}|_{\dot{\alpha}}^I |\mathbf{j}]^{\dot{\alpha} J} \\
    \la\mathbf{i}|p_k|\mathbf{j}\ra = \la{\mathbf{i}}|^{\alpha I} p_{k \alpha\dot{\beta}}|\mathbf{j}\ra^{\dot{\beta}J}, &\quad [\mathbf{i}|p_k|\mathbf{j}\ra = [\mathbf{i}|^I_{\dot{\alpha}}p_k^{\dot{\alpha}\beta} |\mathbf{j}\ra_\beta^J \nn
\end{align}
In this notation, the Dirac equation reads
\bea
\label{eq:dirac}
&&\la\mathbf{i}|p_i = m_i[\mathbf{i}| \quad\quad p_i|\mathbf{i}] = -m_i|\mathbf{i}\ra \\
\nonumber && p_i |\b{i}\ra = -m_i | \b{i} ]     \quad\quad [\b{i}|p_i = m_i \la \b{i} | 
\eea
We can expand $\lambda_\alpha^I$ and $\tilde{\lambda}_{\dot{\alpha}I}$ as explained in eq.(\ref{eq:spinorexpansion})
\begin{align}
    \lambda_\alpha^I &= \lambda_\alpha\, \zeta^{-I} + \eta_\alpha \,\zeta^{+I} \\
                    &= \sqrt{E + p} \,\zeta_\alpha^{+}(p)\, \zeta^{-I}(k) + \sqrt{E-p} \,\zeta_\alpha^-(p)\,\zeta^{+I}(k)\nn\\
    \tilde{\lambda}_{\dot{\alpha}I} &= \tilde{\lambda}_{\dot{\alpha}}\,\zeta_I^+ - \tilde{\eta}_{\dot{\alpha}}\,\zeta_I^- \\
                    &= \sqrt{E+p}\, \tilde{\zeta}_\alpha^-(p)\, \zeta_I^+(k) - \sqrt{E-p} \,\tilde{\zeta}_{\dot{\alpha}}(p) \,\zeta_I^-(k)\nn
\end{align}
where
\bea
&&\zeta_I^+ = \begin{pmatrix} 1\\ 0 \end{pmatrix}\:, \quad \zeta_I^- = \begin{pmatrix} 0 \\-1 \end{pmatrix} \:,  \quad\zeta^{+I} = \begin{pmatrix}0\\1\end{pmatrix} \:,\quad \zeta^{-I} = \begin{pmatrix}1\\0\end{pmatrix}
\eea
\beas
&&\zeta^+_\alpha (p) = \begin{pmatrix} \text{cos }\frac{\theta}{2} \\ \text{sin }\frac{\theta}{2} e^{i \phi} \end{pmatrix} \:, \quad \zeta^-_\alpha (p) = \begin{pmatrix} -\text{sin }\frac{\theta}{2} e^{-i \phi} \\ \text{cos }\frac{\theta}{2} \end{pmatrix} \:\\     &&\zeta^+_{\dot{\alpha}} (p) = \begin{pmatrix} -\text{sin }\frac{\theta}{2} e^{i \phi}  \\ \text{cos }\frac{\theta}{2} \end{pmatrix} \:, \quad \zeta^-_{\dot{\alpha}} (p) = \begin{pmatrix} \text{cos }\frac{\theta}{2} \\ \text{sin }\frac{\theta}{2} e^{-i \phi} \end{pmatrix} 
\eeas
With this choice, we have have the following contractions
\begin{eqnarray}
&&\lambda^I_\alpha \zeta^{+}_I = \lambda_\alpha \qquad   \lambda^I_\alpha \zeta^{-}_I = -\eta_\alpha \nn\\
&&\tilde{\lambda}^I_{\dot{\alpha}} \zeta^{+}_I = -\tilde{\eta}_{\dot{\alpha}} \qquad \tilde{\lambda}^I_{\dot{\alpha}} \zeta^{-}_I = -\tilde{\lambda}_{\dot{\alpha}}
\end{eqnarray}
Furthermore, using \ref{eq:dirac} we can deduce following relations between $\lambda$ and $\eta$
\begin{eqnarray}
&& p_{\alpha \dot{\alpha}}\lambda^{\alpha}=-m\, \tilde{\eta}_{\dot{\alpha}} \qquad p_{\alpha \dot{\alpha}}\eta^{\alpha}=m\,\tilde{\lambda}_{\dot{\alpha}}\nn\\
&& p_{\alpha \dot{\alpha}}\tilde{\eta}^{\dot{\alpha}}=m\,{\lambda}_{\alpha} \qquad p_{\alpha \dot{\alpha}}\tilde{\lambda}^{\dot{\alpha}}=-m\,\eta_{\alpha}
\end{eqnarray}

\section{Amplitudes with one massless particle and 2 equal mass particles}
\label{sec:mincoupling}
Three particle amplitudes involving one massless particle and two massive particles of equal mass present a difficulty. Consider the three particle amplitude with both particles 1 and 2 having mass $m$, spins $S_1$ and $S_2$ and a third  massless particle of helicity h. In order to construct such amplitudes, it is useful to have a Lorentz invariant object which has the correct helicity weight for particle three and is invariant under the little groups for particles 1 and 2. Unfortunately, the obvious candidates vanish,
\bea
\label{eq:degenerate}
[3| \, p_1 \, |3\rangle =2\, p_1. p_3 = 0 \qquad  [3| \, p_2 \, |3\rangle = 2 \, p_2.p_3 = 0 \,.
\eea 
The $x$-factors defined in \citep{massivesh} solve this problem. In our paper, we adopt a slightly different definition and notation which we explain below. For all outgoing momenta, we define
\bea
\label{eq:xfactordef}
   \frac{(p_1 - p_2)_{\alpha\dot\alpha}}{2m} \lambda_3^\alpha = x_{12}^+\,\, \tilde\lambda_{3\dot\alpha} \qquad 
   \frac{(p_1 - p_2)_{\alpha\dot\alpha}}{2m} \tilde\lambda_3^{\dot\alpha} = x_{12}^-\,\, \lambda_{3\alpha} \, .
\eea
Under little group scaling of particle 3, the helicity spinors scale as $\lambda_3 \rightarrow t^{-1} \lambda_3$ and $\tilde{\lambda}_3 \rightarrow t \, \tilde{\lambda}_3$. It follows that $x^+_{12} \rightarrow t^{-2}\, x^+_{12}$ and $x^-_{12} \rightarrow t^{2} \, x^-_{12}$. An object with helicity $h$ transforms as $t^{-2h}$ under a little group scaling. This justifies the $\pm$ signs on the $x$-factors.\\

\ndt We can obtain explicit expressions for the $x$-factors by contracting eq.(\ref{eq:xfactordef}) with reference spinors $\xi^\alpha$ or $\tilde{\xi}^{\dot{\alpha}}$.
\beas
x^+_{12} = \frac{\langle 3|p_1 - p_2 |\xi]}{2 \, m [3\,\xi]} \qquad x^-_{12} = \frac{\langle \xi|p_1 - p_2 |3]}{2 \, m \la \xi \, 3\ra}
\eeas  
These are the same as the conventional expressions for polarization vectors of massless particles upto a factor of $\frac{1}{m}$. It is crucial that the $x$-factors are independent of the reference spinor. To see this, consider two different definitions of $x^+_{12}$  with reference spinors $\xi_1$ and $\xi_2$. Their difference,
\beas
 \frac{\langle 3|p_1 - p_2 |\xi_1]}{2 \, m [3\,\xi_1]} - \frac{\langle 3|p_1 - p_2 |\xi_2]}{2 \, m [3\,\xi_2]} = -\frac{\langle 3|p_1 - p_2 |3][\xi_1 \, \xi_2]}{2 \, m [3\,\xi_1][3\,\xi_2]} = 0
\eeas 
where the first equality follows from a Sch$\ddot{\text{o}}$uten identity and the second from eq.(\ref{eq:degenerate}).\\

\ndt We can build three point amplitudes using the $x$-factors. Here, we will focus on the amplitude involving two spin 1 particles of mass $m$ and a massless particle of helicity $\pm 1$. The contributing structures are
\bea
\label{eq:structures}
\langle \b{12} \rangle^2 \,  x^{\pm}_{12} \qquad [\b{12}]^2 \, x^{\pm}_{12} \qquad \la \b 1 3\ra \, [\b 2 3] \la \b{12}\ra \qquad \la \b 1 3\ra \, \la \b 2 3 \ra \la \b{12}\ra\qquad \dots\, 
\eea  
We pick our amplitudes to be $\la \b{12} \ra^2\, x^-_{12}$ and $[ \b{12} ]^2\, x^+_{12}$. This corresponds to minimal coupling. For more details about this and amplitudes corresponding to multipole moments, see \cite{applications4}. We can also compare these with the vertices that we get from the usual Feynman rules (for a photon with positive helicity).
\begin{equation}
\label{eq:feynman3pt}
    \epsilon_3^+ \cdot (p_1-p_2)\, \epsilon_1 \cdot \epsilon_2 \,+\, \epsilon_1 \cdot (p_2-p_3)\, \epsilon_2 \cdot\epsilon_3^+ \,+\, \epsilon_2 \cdot (p_3-p_1) \, \epsilon_3^+\cdot\epsilon_1
\end{equation}
where 
\beas
(\epsilon_3^+)_{\alpha \dot{\alpha}} \equiv  \, \frac{\lambda_{3\alpha}\, \xi_{\dot{\alpha}}}{[3\, \xi]}\,  \qquad (\epsilon_1)_{\alpha \dot{\alpha}}^{I_1\,I_2} \equiv \frac{1}{m} \, \lambda_{1\alpha}^{\{ I_1} \tilde{\lambda}_{1\dot{\alpha}}^{I_2\}} \qquad (\epsilon_2)_{\beta \dot{\beta}}^{J_1\,J_2} \equiv \frac{1}{m} \, \lambda_{2\beta}^{\{ J_1} \tilde{\lambda}_{2\dot{\beta}}^{J_2\}}
\eeas 
Using these definitions in eq.(\ref{eq:feynman3pt}) and applying Sch$\ddot{\text{o}}$uten identities to eliminate the reference spinors, it reduces to  
\begin{equation}
    \frac{ \, x^+_{12}}{2m} [\b{12}]\left( \la\b{21}\ra + \frac{\la\b{1}3\ra \la3\b{1}\ra}{m \, x^+_{12}} \right) =   -\frac{ \, x^+_{12}}{2m} [\b{12}]^2
\end{equation}
The following identities are useful in showing this equality. 
\begin{eqnarray}
\label{eq:useful}
    [\b{12}] && = \la\b{12}\ra - \frac{\la\b{1}|P|\b{2}]}{m}\\
             && = \la\b{12}\ra +\frac{1}{2m}\left( [\b{1}|P|\b{2}\ra-\la \b{1}|P|\b{2}]\right)\nn
\end{eqnarray}
\begin{equation}
    [\b{21}] = \la \b{21} \ra + \frac{\la \b{2}3\ra \la 3 \b{1}\ra}{m x^+} = \la \b{21} \ra + \frac{[\b{2}3][3\b{1}]}{m x^-}
\end{equation}
Similarly, for a negative helicity photon we have
\begin{equation}
    \frac{- \, x^-_{12}}{2m}\la\b{21}\ra \left( [\b{12}] + \frac{[\b{2}3][3\b{1}]}{mx^-} \right) =   \frac{\, x^-_{12}}{2m}\la\b{21}\ra^2
\end{equation}
\section{Computation of 4-particle amplitudes}
\label{sec:gluing}
In Sections [\ref{sec:wwscattering}] and [\ref{sec:WZ scattering}], we need to glue together two three point amplitudes to construct the four point amplitude. In cases in which the exchanged particle has spin 1, the following identities are useful. 
\begin{align}
    \epsilon_{I_1 J_1} \epsilon_{I_2J_2} \b{I}^{\alpha \{ I_1} \tilde{\b{I}}^{\dot{\alpha}I_2\}}\b{I}^{\beta \{J_1} \tilde{\b{I}}^{\dot{\beta}j_2\}} &= \frac{1}{2}\left(\epsilon^{\alpha\beta}\epsilon^{\dot{\alpha}\dot{\beta}} m_I^2 - I^{\dot{\alpha}\beta} I^{\dot{\beta}\alpha} \right) = \epsilon^{\alpha\beta}\epsilon^{\dot{\alpha}\dot{\beta}} m_I^2 - \frac{1}{2} I^{\dot{\alpha}\alpha}I^{\dot{\beta}\beta}
\end{align}
Note that the second equality can be obtained by using a property of the two dimensional Levi Civita tensor.
\begin{equation}
  \epsilon_{I_1J_1}\epsilon_{I_2J_2} + \epsilon_{I_1I_2}\epsilon_{J_2J_1} + \epsilon_{I_1J_2}\epsilon_{J_1I_2} = 0 
\end{equation}
The following identities are useful in the computation of the 4 point amplitude in Section [\ref{sec:schannel}].
\begin{alignat}{2}
    \b{[12]} &=\b{\la12\ra }- \frac{\b{\la1}I\ra\la I\b{2}\ra}{mx_{12}^+} & &= \b{\la 12\ra} - \frac{[\b{1}I][I\b{2}]}{mx_{12}^-} \\
    \b{\la 34\ra} &= \b{[34]} + \frac{\la \b{3}I \ra\la I \b{4}\ra}{m_{34}^+} & &= [\b{34}] + \frac{[\b{3}I][I\b{4}]}{m x_{34}^-}
\end{alignat}
and
\begin{align}
    x_{12}^+x_{34}^- + x_{12}^-x_{34}^+ &= \frac{1}{2m^2}(p_1 - p_2).(p_3-p_4)
\end{align}

\section{Generators of $SO(4)$ and the embedding of $SU(2)\times U(1)_Y$}
\label{sec:grouptheory}
The generators of SO(4) are
\beas
&& A_1 = i\left(
\begin{array}{cccc}
 0 & 0 & 0 & 0 \\
 0 & 0 & -1 & 0 \\
 0 & 1 & 0 & 0 \\
 0 & 0 & 0 & 0 \\
\end{array}
\right) \:, \quad
A_2 = i\left(
\begin{array}{cccc}
 0 & 0 & 1 & 0 \\
 0 & 0 & 0 & 0 \\
 -1 & 0 & 0 & 0 \\
 0 & 0 & 0 & 0 \\
\end{array}
\right) \:, \quad
A_3 = i\left(
\begin{array}{cccc}
 0 & -1 & 0 & 0 \\
 1 & 0 & 0 & 0 \\
 0 & 0 & 0 & 0 \\
 0 & 0 & 0 & 0 \\
\end{array}
\right)\\
&& B_1 = i\left(
\begin{array}{cccc}
 0 & 0 & 0 & -1 \\
 0 & 0 & 0 & 0 \\
 0 & 0 & 0 & 0 \\
 1 & 0 & 0 & 0 \\
\end{array}
\right) \:, \quad
B_2 = i\left(
\begin{array}{cccc}
 0 & 0 & 0 & 0 \\
 0 & 0 & 0 & -1 \\
 0 & 0 & 0 & 0 \\
 0 & 1 & 0 & 0 \\
\end{array}
\right) \:, \quad
B_3 = i\left(
\begin{array}{cccc}
 0 & 0 & 0 & 0 \\
 0 & 0 & 0 & 0 \\
 0 & 0 & 0 & -1 \\
 0 & 0 & 1 & 0 \\
\end{array}
\right)
\eeas
The combinations
\beas
&& X^+ = \frac{1}{2\sqrt{2}}\left( A_1+A_2 + i (B_1+ B_2)\right)  \quad X^- = \frac{1}{2\sqrt{2}}\left( A_1+A_2 - i (B_1+ B_2)\right) \quad  X^3 = \frac{1}{2} \left( A_3 + B_3\right)\\
&& Y^+ = \frac{1}{2\sqrt{2}}\left( A_1-A_2 + i (B_1 - B_2)\right)  \quad Y^- = \frac{1}{2\sqrt{2}}\left( A_1-A_2 - i (B_1- B_2)\right) \quad  Y^3 = \frac{1}{2} \left( A_3 - B_3\right)
\eeas
satisfy two copies of $SU(2)$, i.e
\begin{alignat}{3}
   \left[ X^+, X^-\right] &= 2X^3 ,\quad \left[ X^+, X^3\right] &= X^+, \quad \left[ X^-, X^3\right] &= -X^-, \nonumber\\
\left[ Y^+, Y^-\right] &= 2Y^3 ,\quad \left[ Y^+, Y^3\right] &= Y^+, \quad \left[ Y^-, Y^3\right] &= -Y^- ,
\end{alignat}

We will associate the generators $X^\pm,X^3$ with the symmetry $SU(2)_L$. These are referred to as $T^\pm, T^3$ in the paper. The $U(1)_Y$ is a subgroup of the $SU(2)$ formed by $Y^\pm, Y^3$ and we will set $Y^3 \equiv T^B$.

\bibliography{sample}
\bibliographystyle{unsrt}

\end{document}